\documentstyle[epsfig, subfigure, lscape]{mn}

\def\ltapprox{\raise 2pt \hbox {$<$} \kern-1.1em \lower 5pt \hbox {$\approx$}}
\def\ltsim{\raise 2pt \hbox {$<$} \kern-1.1em \lower 4pt \hbox {$\sim$}}
\def\gtsim{\raise 2pt \hbox {$>$} \kern-1.1em \lower 4pt \hbox {$\sim$}}

\title{Alfv\'enic reacceleration of relativistic particles in galaxy clusters
in the presence of secondary electrons and positrons}
\author[G. Brunetti \& P. Blasi]
      {G. Brunetti,$^1$ 
       P. Blasi$^2$ \\
       $^1$ INAF/Istituto di Radioastronomia, via Gobetti 101,
       I--40129 Bologna, Italy \\
       $^2$ INAF/Osservatorio Astrofisico di Arcetri,
       Largo E. Fermi 5, I-50125 Firenze, Italy\\
}

\begin{document}
\maketitle

\begin{abstract}
In a previous paper (Brunetti et al. 2004) we presented the first 
self-consistent calculations of the time-dependent coupled equations
for the electrons, hadrons and Alfv\'en waves in the intracluster
medium, which describe the stochastic acceleration of the charged 
particles and the corresponding spectral modification of the waves. 
Under viable assumptions, 
this system of mutually interacting components was shown to accurately
describe several observational findings related to the radio halos in 
clusters of galaxies. 

In this paper, we add to the self-consistency of the calculations by 
including the generation and re-energization of secondary electrons
and positrons, 
produced by the inelastic interactions of cosmic rays with the 
thermal gas in the intracluster medium. The bulk of Cosmic rays is 
expected to be confined within the cluster volume for cosmological 
times, so that the rate of production of secondary electrons, as
well as gamma rays, may become correspondingly enhanced. If MHD waves 
are present, as it may be expected in the case of a recent merger
event, then the reacceleration of secondary electrons and positrons
can
significantly affect the phenomenology of the non thermal processes 
in clusters. We investigate here these effects for the first time.
\end{abstract}


\maketitle

\section{Introduction}

The intracluster medium contains a non thermal component in the
form of magnetic fields and relativistic particles, as demonstrated
mainly by the appearance of diffuse radio emission (e.g., Feretti, 2003)
and by studies of the rotation measure of radio sources in galaxy
clusters (e.g., Clarke et al., 2001),
but also by the hard X-ray (HXR) excess detected in
a few galaxy clusters by the BeppoSAX and RXTE satellites
(Fusco-Femiano et al.~2003; Rephaeli \& Gruber ~2003). While the 
radio radiation is certainly the result of synchrotron emission
of high energy electrons in the intracluster magnetic field,
HXRs may be explained in terms of inverse Compton scattering (ICS) 
of the same electrons off the photons of the cosmic microwave background
(CMB), or as a result of bremsstrahlung emission from supra-thermal
electrons (e.g., Ensslin, Lieu, Biermann 1999; Blasi 2000; Dogiel 2000; 
Sarazin \& Kempner 2000). The latter solution does however require an
energy input which may be supported only for a time shorter than 
a few hundred million years, otherwise the gas is overheated 
(Petrosian 2001; Blasi 2000). It should be mentioned that 
the rather poor sensitivity of the present and past facilities 
for the observation of HXRs does not allow to obtain a iron--clad 
detection of HXR excesses. Future observatories (e.g. ASTRO-E2, NEXT) 
are necessary to confirm (or disprove) the existence of these 
excesses (see Rossetti \& Molendi 2004; Fusco-Femiano et al.~2004). 

So far the most serious problem from the theoretical point of view is
to identify the origin of the radiating high energy electrons.
Buote (2001) and Schuecker et al. (2001) have found hints 
of a correlation between the non thermal diffuse radio emission 
and the presence of merger activity in the host clusters. 
This may indicate a link between the process of formation of
galaxy clusters and the origin of the non thermal activity.

Two main avenues have been identified to explain the fact that
such high energy electrons are present and able to radiate on
distance scales larger than their typical loss lengths: in 
the context of the so-called {\it primary} models 
electrons are accelerated at shock waves through the first order Fermi 
mechanism or they are continuously re-energized {\it in situ} on 
their way out (Jaffe 1977); the latter case is also called
{\it reacceleration} model.
On the other hand, in the context of {\it secondary electron} models
electrons are secondary products of the hadronic interactions of 
cosmic rays with the intracluster medium, as first proposed by Dennison (1980).

Shock waves are unavoidably formed during merger events due to
the supersonic relative motion of two (or more) infalling clusters.
Since electrons have a short pathlength due to IC losses, they can
only move a short distance away from the acceleration regions.
If this acceleration site were to be at a shock front, the emission
from the electrons would be concentrated around the shock rim 
(e.g., Miniati et al. 2001) and
the spectrum of the radiation would be quite steep because of 
the low Mach numbers in the central virialized regions of clusters
(Gabici and Blasi 2003, Berrington and Dermer 2003), where the radio
emission is observed to come from. On the other hand, strong shocks
are formed in the outskirts of clusters, and they may be responsible
for the acceleration of protons (Ryu et al. 2003). These protons can 
then be advected into the cluster and there be confined for cosmological 
time scales (V\"{o}lk et al. 1996; Berezinsky, Blasi \& Ptuskin 1997).
The confinement unavoidably increases the energy density of cosmic 
rays in the intracluster medium, and may correspondingly favour
the generation of gamma rays through the decay of neutral pions
and of electrons and positrons through the decay of charged pions. 

The origin of the emitting particles is still matter of
debate (e.g., Ensslin 2004). The different models for the production of 
the radiating electrons have however a substantial predictive power,
which can be used to discriminate among such models by comparing their 
predictions with observations.
Recently Brunetti (2004) and Blasi (2004)
have discussed at length the strong and the weak points of the models 
in explaining the phenomenology of existing data. 
Although the role of future observations remains crucial in order
to achieve a definite conclusion of the right descriptions of the non thermal
phenomena we detect, both authors concluded 
that, at least as far as the Coma cluster and few other well 
studied clusters are concerned, present 
data seem to suggest the presence of particle--reacceleration
mechanisms active in the ICM. On the
other hand, several complex and poorly understood processes are 
involved in these models.

Cluster mergers induce large--scale bulk flows with velocities 
$\sim 1000$ km s$^{-1}$ or larger. These flows drive instabilities on 
large--scales which redistribute the energy of the mergers through 
the cluster volume and decay into turbulent velocity fields.
It has been shown that reacceleration of a population of relic 
electrons by turbulence powered by cluster mergers is a promising
mechanism to explain the very large scale of the observed radio 
emission, the complex spectral behaviour observed in some diffuse 
radio sources (Brunetti et al., 2001; Petrosian 2001; Ohno, Takizawa 
and Shibata 2002; Fujita, Takizawa and Sarazin 2003), and the
observed occurrence of radio halos with cluster mass (Cassano \& Brunetti 
2005).

A step forward in the study of particle acceleration in galaxy clusters
has been achieved by recent studies of the interaction between
particles and Alfv\'en waves in a very general situation in which 
relativistic electrons, thermal protons and relativistic protons 
are present in the cluster volume (Brunetti et al., 2004, hereafter 
Paper I). In Paper I the interaction of all these components with 
the waves, as well as the turbulent cascading and damping processes 
of Alfv\'en waves, have been treated in a fully time-dependent way 
in order to calculate the spectra of electrons, protons and waves at 
any fixed time. This represented the first attempt to include in a
self-consistent way hadronic cosmic rays in the reacceleration
scenario. It was found there that radio halos and HXR tails could
be activated for a time scale of $\sim 0.5-1$ Gyr through
resonant interaction of particles with short-wavelength Alfv\'en 
waves, coming from the decay of merger induced turbulence. The role 
of the hadrons was found to be that of exerting a substantial 
backreaction, so that the non thermal activity is suppressed 
if the energy content in the form of hadrons gets larger than
a few percent of the thermal energy. We named this phenomenon
{\it wave-proton boiler}.

The present paper serves as a completion of the effort started in
Paper I: here we add to the previous calculation the effect of the 
reacceleration of those electrons and positrons
that are generated at any time as 
secondary products of the inelastic interactions of the confined 
cosmic rays with the thermal gas.

In Sections 2 and 3 we provide the reader with a review of the main
aspects of the calculations developed in Paper I. In Section 4 we
describe the general scenario in which secondary reaccelerated electrons
are present. The application of the full calculation to the phenomenology
of a Coma-like cluster is presented in Section 5. We conclude in Section 6.

\section{Particle acceleration and energy losses in the ICM}

In this Section we briefly summarize the rates of energy loss and
gain for non thermal leptons and protons in the intracluster 
medium, relying on the formalism already introduced in Paper I.

\subsection{Energy losses for leptons}

Four channels dominate the energy losses of relativistic 
leptons with momentum $p$, namely ionization, Coulomb scattering,
synchrotron emission and IC. The rate of losses due to the 
combination of ionization and Coulomb scattering can be written
following (Sarazin 1999):

\begin{equation}
\left( {{ d p }\over{d t}}\right)_{\rm i} 
=- 3.3 \times 10^{-29} n_{\rm th}
\left[1+ {{ {\rm ln}(\gamma/{n_{\rm th}} ) }\over{
75 }} \right]
\label{ion}
\end{equation}

\noindent
where $n_{\rm th}$ is the number density of the thermal plasma.
The rate of synchrotron plus IC losses is:
\begin{equation}
\left( {{ d p }\over{d t}}\right)_{\rm rad}
=- 4.8 \times 10^{-4} p^2
\left[ \left( {{ B_{\mu G} }\over{
3.2}} \right)^2 {{ \sin^2\theta}\over{2/3}}
+ (1+z)^4 \right]
\label{syn+ic}
\end{equation}

\noindent
where $B_{\mu G}$ is the magnetic field strength in
units of $\mu G$, and $\theta$ is the pitch angle of the emitting 
leptons; in case of efficient isotropization of the 
electron momenta, the $\sin^2\theta$ is averaged to $2/3$.
In the typical conditions of the ICM, radiative losses are 
dominant for leptons with Lorentz factor $\gamma >> 100$, 
while Coulomb losses dominate at lower energies (Sarazin 1999,2002; 
Brunetti 2003).

\subsection{Energy losses for protons}

For relativistic protons, the main channel of energy losses 
in the ICM is provided by inelastic proton-proton collisions.
The time-scale associated with this process is:

\begin{equation}
\tau_{pp} = 
{ 1\over
{{ 
n_{\rm th} \sigma_{\rm pp} c
}} }
\sim 10^{18} \Big(
{{ n_{\rm th} }\over{ 10^{-3}}}
\Big)^{-1}~s.
\label{taupp}
\end{equation}

\noindent
Thus inelastic ${\rm pp}$ scattering is weak enough to
allow the accumulation of protons over cosmological times
(Berezinsky, Blasi and Ptuskin 1997), however it is 
also efficient enough for the continuous production of pions,
which in turn decay into gamma rays (for neutral pions), 
electron-positron pairs and neutrinos (for charged pions). 
The process of pion production in ${\rm pp}$ scattering is
a threshold reaction that requires protons with kinetic energies 
larger than about 300 MeV.

For trans-relativistic and mildly relativistic protons, energy
losses are dominated by ionization and Coulomb scattering. 
Protons more energetics than the thermal electrons, namely with 
$\beta_p > \beta_c \equiv (3/2 m_e/m_p)^{1/2} \beta_e$
($\beta_e \simeq 0.18 (T/10^8 K)^{1/2}$ is the velocity of the 
thermal electrons), are affected by Coulomb interactions.
Defining $x_m \equiv \left( {{ 3 \sqrt{\pi}}\over{4}}
\right)^{1/3}\beta_e$, one has (Schlickeiser, 2002):

\begin{equation}
{{ d p }\over{dt}} \simeq
- 1.7 \times 10^{-29}
\left( {{n_{\rm th}}\over{10^{-3}}} \right)
{{
\beta_p }\over{
x_m^3 + \beta_p^3 }}~\rm cgs~units,
\label{coulomb_p}
\end{equation}

\noindent
which has the following asymptotic behaviour :

\begin{equation}
{{ d p }\over{dt}} 
\propto
\left( {{n_{\rm th}}\over{10^{-3}}} \right)
\times 
\left\lbrace \begin{array}{lll}
\textrm p      & \textrm{for} & mc \beta_c < p < mc x_m \\
p^{-2} & \textrm{for} & mc x_m < p << m c \\
{\textrm{Const.}} & \textrm{for} & p>> m c 
\end{array}
\right.
\label{coulomb_pasy}
\end{equation}

\subsection{Alfv\'enic acceleration of relativistic particles}

Different coupling between particles and waves may result in 
energy transfer from magnetic fluctuations into relativistic
particles: Magneto-Sonic (MS) waves, magnetic Landau damping  
(Kulsrud \& Ferrari 1971; Schlickeiser \& Miller 1998), 
Lower Hybrid (LH) waves (e.g., Eilek \& Weatherall 1999) 
and Alfv\'en waves are a few examples that have been investigated
in the literature. Alfv\'en waves efficiently couple with 
relativistic particles via resonant interaction and they are 
likely to transfer most of their energy directly into these 
relativistic particles.

\noindent
The resonant condition for a wave of frequency $\omega$ and wavenumber 
projected along the magnetic field $k_{\Vert}$, for a particle of 
species $\alpha$ with energy $E_{\alpha}$ and projected velocity
$v_{\Vert}=v \mu$ is (Melrose 1968; Eilek 1979):

\begin{equation}
\omega
- \nu {{\Omega_{\alpha}}\over
{\gamma}} - k_{\Vert} v_{\Vert} =0,
\label{resonance}
\end{equation}

\noindent
where, in the quasi parallel case ($k_{\perp} << m_{\alpha}
\Omega_{\alpha}/p$), $\nu=-1$ ($\nu=1$) for electrons (protons
and positrons).

The dispersion relation for Alfv\'en waves in an isotropic 
plasma with both thermal and relativistic particles was given
by Barnes \& Scargle (1973). In the conditions typical of galaxy 
clusters, the dispersion relation of Alfv\'en  waves reduces to 
$\omega \simeq |k_{\Vert}| v_{\rm A}$. Combining the dispersion 
relation of the waves with the resonant condition, Eq. \ref{resonance}, 
one can derive the resonant wavenumber, $k_{res}$, for a given 
momentum, $p=m v \gamma$, and angle, $\mu$, of the particles :

\begin{equation}
k_{res} \sim | k_{\Vert} | 
= {{ \Omega m}\over{p}}
{ 1 \over
{\Big( \mu \pm {{v_{\rm A}}\over{v}} \Big) }},
\label{kres}
\end{equation}

\noindent
where the upper and lower signs refer to protons (and positrons)
and electrons 
respectively. The interaction of particles with Alfv\'en waves 
can be thought of as a diffusion process in the momentum space of 
the particles. If the distributions of waves and particles are
assumed to be isotropic, then the diffusion coefficient was
found by Eilek \& Henriksen (1984) to be:

\begin{equation}
D_{\rm pp}(p,t) =
{{ 2 \pi^2 e^2 v_{\rm A}^2}\over{c^3}}
\int_{k_{min}}^{k_{max}}
{{ W_k(t)}\over{k}}
\Big[
1 -
\Big(
{{v_{\rm A}}\over{c}} \mp
{{ \Omega m }\over{p k}} \Big)^2
\Big] dk,
\label{dpp}
\end{equation}
\noindent
where the minimum wavenumber (maximum scale length) of the waves 
interacting with particles with given momentum is:
\begin{equation}
k_{min} = {{ \Omega m}\over{p}}
{ 1 \over
{\Big( 1 \pm {{v_{\rm A}}\over{v}} \Big) }},
\label{kres}
\end{equation}
\noindent
and $k_{max}$ is given by the largest wavenumber of the Alfv\'en 
waves, which is fixed by the condition that the frequency of the
waves cannot exceed the proton cyclotron frequency, namely 
$\omega < \Omega_p$. It follows that $k_{max} \sim \Omega_p/v_A$ 
(in the following, for consistency with Paper I, we take $k_{max} 
\sim \Omega_p/v_M$, $v_M$ being the magnetosonic velocity).

\section{Equations and Coupling}

\subsection{Time Evolution of Particles and Waves}

In this Section we summarize the formalism that we introduced
in Paper I for the description of the time-dependent interaction
between the particles and the waves. The evolution of the electron 
(and positron) 
number density is given by the diffusion equation, which includes 
energy losses and gains (Paper I and references therein):

\begin{eqnarray}
{{\partial N_e(p,t)}\over{\partial t}}=
{{\partial }\over{\partial p}}
\left[
N_e(p,t)\left(
\left|{{dp}\over{dt}}_{\rm rad}\right| + \left|{{dp}\over{dt}}_{\rm i}
\right|
-{1\over{p^2}}{{\partial }\over{\partial p}}(p^2 D_{\rm pp})
\right)\right] 
\nonumber\\
+ {{\partial^2 }\over{\partial p^2}}
\left[
D_{\rm pp} N_e(p,t) \right] +
Q_e[p,t;N_p(p,t)].
\label{elettroni}
\end{eqnarray}
\noindent
Here $Q_e[p,t;N_p(p,t)]$ represents the injection rate of 
secondary relativistic electrons and positrons
generated during the collisions 
between the accelerated relativistic protons with the thermal 
protons in the ICM (Sect. 3.3). A similar equation can be written
for protons (Paper I and references therein) :

\begin{eqnarray}
{{\partial N_p(p,t)}\over{\partial t}}=
{{\partial }\over{\partial p}}
\left[
N_p(p,t)\left( \left|{{dp}\over{dt}}_{\rm i}\right|
-{1\over{p^2}}{{\partial }\over{\partial p}}(p^2 D_{\rm pp})
\right)\right] 
\nonumber\\
+ {{\partial^2 }\over{\partial p^2}}
\left[
D_{\rm pp} N_p(p,t)
\right]. 
\label{protoni}
\end{eqnarray}

The evolution of the spectrum of the Alfv\'en  waves is described 
through a diffusion equation in the wavenumber space (Eilek 1979):

\begin{eqnarray}
{{\partial W_{\rm k}(t)}\over
{\partial t}} =
{{\partial}\over{\partial k}}
\left( D_{\rm kk} {{\partial W_{\rm k}(t)}\over{\partial k}}
\right)
-\sum_{i=1}^n \Gamma^{\rm i}(k) W_{\rm k}(t) 
\nonumber\\
+ I_{\rm k}(t),
\label{turbulence}
\end{eqnarray}
\noindent
where $D_{\rm kk}$ is the diffusion coefficient due to wave-wave
coupling (Sect. 3.2.2), $\Gamma$ is the damping rate of the Alfv\'en 
waves with particles (Sect. 3.2.1) and $I_k$ is the injection rate 
of the Alfv\'en waves (Sect. 3.2.3).
In Eq.(\ref{turbulence}) we use the assumption, commonly
made, that wave--wave interaction is just local in the wave
number space.

\subsection{Damping Procesess, wave-wave coupling
and injection of turbulence}

All the relevant processes related to wave-particle interactions, 
wave-wave interactions and injection of Alfv\'en waves in the ICM 
are described in detail in Paper I. Here we only provide the reader 
with a brief overview of the main processes that we include in our 
calculations.

\subsubsection{Damping Processes}

In the case of nearly parallel wave propagation
(i.e., $k_{\perp} << m \Omega/p$, $k\simeq k_{\Vert}$) 
and isotropic distribution of the velocities of the particles,
the cyclotron resonant damping rates for Alfv\'en  waves with 
particles of species $\alpha$ are given by Melrose (1968):

\begin{eqnarray}
\Gamma^{\alpha}_{\rm k}(t)=
-{{ 4 \pi^3 e^2 v_{\rm A}^2 }\over
{k c^2}}
\int_{p_{\rm min}}^{p_{\rm max}}
p^2 (1 - \mu_{\alpha}^2 )
{{ \partial f_{\alpha}(p,t) }\over{ \partial p}}
dp =\nonumber\\
{{ \pi^2 e^2 v_{\rm A}^2 }\over
{k c^2}}
\int_{p_{\rm min}}^{p_{\rm max}}
(1 - \mu_{\alpha}^2 )
\left(
2 {{N_{\alpha}(p,t)}\over{p}}
-{{\partial N_{\alpha}(p,t) }\over{ \partial p}}
\right) dp,
\label{damping}
\end{eqnarray}
\noindent
where, in the relativistic case one has :
\begin{equation}
\mu_{\alpha}^{\rm rel}=
{{ v_{\rm A} }\over
{c}} \pm {{\Omega_{\alpha} {\rm m}_{\alpha} }\over
{p k}}.
\label{damping_1}
\end{equation}
\noindent
and in the non relativistic case:
\begin{equation}
\mu_{\alpha}^{\rm th}=
{{v_{\rm A} {\rm m}_{\alpha} }\over
{p}}
\pm
{{\Omega_{\alpha} {\rm m}_{\alpha} }\over
{p k}}.
\label{damping_2}
\end{equation}
\noindent
The upper and lower signs in Eqs. \ref{damping_1}-\ref{damping_2}
are for negative and positive charged particles respectively. 
Eq. \ref{damping} can also be used to evaluate the damping rate
in the case of isotropic Alfv\'en waves with an approximation
which is within a factor of $\sim 3$ (Lacombe 1977).

\subsubsection{Wave-wave Cascade}

Wave-wave interactions cause the spectrum of the waves to cascade,
namely to broaden toward larger values of $k$. This is a diffusive
process, with diffusion coefficient $D_{\rm kk} = k^2/\tau_s$. The
time $\tau_s$ is the spectral energy transfer time and can be written 
as $\tau_s \sim \tau_{NL}^2/\tau_3$ (Zhou \& Matthaeus 1990), where 
$\tau_{NL}=\lambda/\delta v$ is the non-linear eddy-turnover time
($\delta v$ is the rms velocity fluctuation at wavelength $\lambda$) 
and $\tau_3$ is the time over which this fluctuation interacts with 
other fluctuations of similar size.

In the context of the Kolmogorov phenomenology, the Alfv\'en  crossing 
time $\tau_{A}=\lambda/v_A$ exceeds $\tau_{NL}$ and the fluctuations
of comparable size interact in one turnover time, namely $\tau_3 \sim 
\tau_{NL}$. Since the velocity fluctuation, $\delta v$, is related to 
the rms wave field, $\delta B$, by $\delta v^2 / v_A^2 = \delta B^2 /B^2$, 
the diffusion coefficient can be written as (Miller \& Roberts 1995):

\begin{equation}
D_{\rm kk} \simeq v_{\rm A} 
k^{7/2}
\left( {{ W_{\rm k}(t) }\over{ 2 W_{\rm B}}} \right)^{1/2}.
\label{dkk}
\end{equation}
\noindent
Given a spectrum of injection of waves per unit time, $I_k$, one simple 
possibility to estimate the cascade time scale is to use the spectum of 
the waves in Eq. \ref{dkk} as obtained from Eq. \ref{turbulence}
under stationary conditions and without damping processes. In Paper I
we found:
\begin{equation}
\tau_s = {{k^2}\over{D_{kk}}}
\sim
{1\over k}
\Big(
{{B^2}\over{4 \pi}}
\Big)^{{1\over3}} 
\Big( v_A^2  I_k \Big)^{-{1\over 3}}.
\label{tau-ww}
\end{equation}

\subsubsection{Injection of Alfv\'en waves in the ICM: the
{\it Lighthill} mechanism}

While the physics involved in the process of energy transfer
between waves and particles for a given spectrum of waves 
is relatively well understood, the transformation of the wave
spectrum starting from some injection at large scales is 
rather poorly known. The waves are expected to couple with 
relativistic particles when the turbulence has been enriched 
of short wavelength modes, so that the cascading is implicitely
required to be rather efficient if the injection occurs on 
macroscopic scales. If however this is the case, it was shown
(Yan \& Lazarian 2004 and refs. therein) that the Alfv\'en waves
reach the high-k part of the spectrum with a highly anisotropic 
spectrum, and the efficiency of particle acceleration is likely
to be therefore drastically reduced. From this follows that the
acceleration process is favored in those scenarios in which the
injection of Alfv\'en waves occurs on relatively small scales
to start with. One injection process in which this condition
is fulfilled is provided by the so-called {\it Lighthill} mechanism 
(Kato 1968; Eilek \& Henriksen 1984), which can convert some
fraction of the large scale fluid turbulence on the larger scales
into Alfv\'en waves on smaller scales. Following Fujita et al. (2003)
and Paper I, we assume in our calculation that fluid turbulence
is injected on large scales, for instance excited by a merger 
event, and that the {\it Lighthill} mechanism couples the fluid turbulence
with MHD turbulence on smaller scales. 
We made the assumption here that the spectrum of the fluid turbulence
(not the MHD turbulence) is in the form of a power law
\begin{equation}
W_{\rm f}(x_{\rm f})=
W_{\rm f}^{\rm o} x_{\rm f}^{-{\rm m}},
\label{wfluid}
\end{equation}
\noindent
in the range $x_{\rm o} < x_{\rm f} < x_{\rm f}^{\rm max}$,
where $x_{\rm o}$ is the wavenumber corresponding to the maximum 
scale of injection of the turbulence; we thus do not consider
the possibility that turbulence may be injected simultaneously
at many scales (e.g., Tsytovich 1972).
The maximum wavenumber is where 
the effect of fluid viscosity starts to be important and it is 
of the order of $x_{\rm f}^{\rm max} \sim x_{\rm o} 
( {\cal R} )^{3/4}$ (Landau \& Lifshitz, 1959), ${\cal R}$ being the 
Reynolds' number. Thus for high values of the Reynolds number in the
ICM (and in a magnetized medium) the turbulence cascading can be an
efficient process down to scales much smaller than kpc (Fujita et al.
2003; Paper I).

In the {\it Lighthill} process a fluid eddy may be thought of as 
radiating Alfv\'en waves at a wavenumber $k = (v_{\rm f}(x)/v_A)x_{\rm f}$.
The Alfv\'en waves are expected to be driven only for $x_f > x_{\rm T}$, 
$x_{\rm T}$ being the wavenumber at which the transition
from large-scale ordered turbulence to small-scale disordered turbulence
occurs. This transition is usually assumed to take place at the Taylor 
scale (Eilek \& Henriksen 1984), 
$l_{\rm T} \sim l_o (15 / {\cal R} )^{1/2}$, where 
the Reynolds number is given by ${\cal R} =l_o v_{\rm f}/\nu_{\rm K}$,
and $\nu_{\rm K}$ is the kinetic viscosity. 

The fraction of the fluid turbulence radiated in the form of MHD modes 
is small for all but the larger eddies, near the Taylor scale. Therefore 
the {\it Lighthill} radiation may be expected to not disrupt the fluid 
spectrum. The rate of radiation {\it via} the {\it Lighthill} mechanism 
into  Alfv\'en waves of wavenumber $k$ is (Eilek \& Henriksen 1984;
Fujita et al.2003; Paper I):

\begin{equation}
I_k
\simeq 2
\Big|
{{ 3 -2 m}\over{3 -m }}
\Big|
\rho
v_{\rm A}^3 
\Big(
{{ v_{\rm f}^2 }\over
{  v_{\rm A}^2
R }}
\Big)^{ 3 \over{3 -m }}
\times 
k^{-3 {{m-1}\over{3-m}}},
\label{inj_alfven}
\end{equation}
\noindent
where $\rho \sim {\cal E}_t / v_f^2$, with ${\cal E}_t$ the energy 
density of the fluid turbulence, and
\begin{equation}
R=
{{ x_o W_f(x_o)}\over
{x_T W_f(x_T)}}.
\label{r}
\end{equation}

\subsection{Secondary Electrons}

As discussed above, the main new ingredient added in this paper,
compared with the calculations presented in Paper I is the 
presence of secondary electrons (and positrons), as generated 
in the hadronic inelastic interactions of cosmic rays with the
thermal gas in the ICM. 
The decay chain that we consider is (Blasi \& Colafrancesco 1999):

$$p+p \to \pi^0 + \pi^+ + \pi^- + \rm{anything}$$
$$\pi^0 \to \gamma \gamma$$
$$\pi^\pm \to \mu + \nu_\mu ~~~ \mu^\pm\to e^\pm \nu_\mu \nu_e.$$

\noindent
The spectrum of secondary electrons and positrons
with energy $E_e$ is given by the 
convolution of the spectra of protons, $N(E_p)$, with the spectrum of 
pions produced in a single cosmic ray interaction at energy $E_p$, 
($F_{\pi} (E_{\pi},E_p)$) and with the distribution of leptons 
from the pion decay, $F_e^{\pm}(E_e,E_{\pi})$, (e.g., Moskalenko \& 
Strong, 1998):

\begin{eqnarray}
Q_e^{\pm}[p,t;N_p] = 
n^p_{th} c 
\int_{E_{tr}} dE_p \beta_p N(E_p) \sigma^{\pm}_{\pi}(E_p)
\nonumber\\
\int dE_{\pi} F_{\pi}(E_{\pi},E_p) F_e^{\pm}(E_e,E_{\pi}),
\label{qepm0}
\end{eqnarray}

\noindent
where $\sigma^{\pm}(E_p)$ is the inclusive cross section for pion 
production, $E_{tr}$ is the threshold energy for the process to occur
and the distribution of electrons and positrons is given by :

\begin{equation}
F_e^{\pm}(E_e,E_{\pi^{\pm}}) =
\int dE_{\mu} F_e^{\pm}(E_e,E_{\mu},E_{\pi}),
\label{feeepi}
\end{equation}

\noindent
where $F_e^{\pm}(E_e,E_\mu,E_\pi)$ is the spectrum of electrons/positrons 
from the 
decay of a muon of energy $E_\mu$ produced in the decay of a pion with 
energy $E_\pi$.

At large values of $E_p$ the differential cross section is
sufficiently well described by the so-called Feynman scaling, with 
small deviations which can easily be taken into account. In the low 
energy part, when the reaction occurs close to the threshold, and in 
general at laboratory energies smaller than $\sim 10$ GeV, the experimental 
data on pion production are rather poor, and the scaling behaviour
is violated. 
Since in this paper we are going to calculate the spectrum of the
reaccelerated secondary electrons and positrons, 
we are forced to use a source term
which correctly describes the spectrum of the injected leptons over 
a broad energy range ($\gamma \sim 10^2-10^5$).
A practical and useful approach to both the high energy and low
energy regimes was proposed in Dermer (1986a) and reviewed by  
Moskalenko \& Strong (1998), and is based on the combination of the
isobaric model (Stecker 1970) and scaling model
(Badhwar et al., 1977; Stephens \& Badhwar 1981). Here we briefly describe
the formalism and approximations used in our calculations and
provide the main equations.

In the Stecker's model the
pion production due to $pp$ collisions near threshold is
mediated by the excitation of the $\Delta_{3/2}$ isobar, which
subsequantly decays into a nucleon and a pion.
In this case 
the spectrum of pions produced in a single cosmic ray interaction
is given by (e.g., Dermer 1986a; Strong \& Moskalenko 1998):

\begin{eqnarray}
F_\pi(E_\pi,E_p)=
\Gamma \int_{m_pc^2 +m_{\pi}c^2}^{\sqrt{s} -m_pc^2}
{{ dm_{\Delta}c^2 f_{\pi}(E_{\pi},E_p; m_{\Delta}) }\over{
(m_{\Delta}c^2 - m_{\Delta}^oc^2)^2 +\Gamma^2}} \times
\nonumber\\
\times 
\Big(
{\rm tan}^{-1}\left(
{{\sqrt{s} -m_pc^2 -m_{\Delta}^oc^2 }\over{\Gamma}}\right)
-
\nonumber\\
{\rm tan}^{-1}\left(
{{m_pc^2 +m_{\pi}c^2-m_{\Delta}^oc^2 }\over{\Gamma}}\right)
\Big)^{-1},
\label{Fpi_steck}
\end{eqnarray}

\noindent
where $\Gamma \simeq 0.0575$ GeV is the width of the Breit--Wigner 
distribution, $m_{\Delta}^oc^2 \simeq 1.232$ GeV is the average rest 
energy of the $\Delta$--isobar, and $s = 2 m_pc^2 (E_p + m_pc^2)$ is 
the square of the energy in the center of mass frame, 

\begin{equation}
f_{\pi}(E_{\pi},E_p; m_{\Delta})=
{1 \over{ 4 m_{\pi}c^2 \gamma_{\pi}^{\prime}
\beta_{\pi}^{\prime}}}
\left(
{{ H^+ }\over{\beta_{\Delta}^+ \gamma_{\Delta}^+ }}
+
{{ H^- }\over{\beta_{\Delta}^- \gamma_{\Delta}^- }}
\right),
\label{f_steck}
\end{equation}

\noindent
where $H^{\pm}=1$ for $ \gamma_{\Delta}^{\pm}
\gamma_{\pi}^{\prime}
(1 - 
\beta_{\Delta}^{\pm}
\beta_{\pi}^{\prime}) \leq \gamma_{\pi} \leq
\gamma_{\pi}^{\prime}
(1 +
\beta_{\Delta}^{\pm}
\beta_{\pi}^{\prime})$ and
$H^{\pm}=0$ otherwise, 

\begin{equation}
\gamma_{\Delta}^{\pm}=
\gamma_c \gamma_{\Delta}^*
(1 \pm \beta_c \beta_{\Delta}^*),
\label{gammadelta}
\end{equation}

\noindent
with 

\begin{equation}
\gamma_{\Delta}^*=
{{ s + m_{\Delta}^2c^4 -m_p^2c^4}\over{
2 \sqrt{s} m_{\Delta} c^2}}
\label{gammadelta*}
\end{equation}

\noindent
the Lorentz factor of the $\Delta$--isobar in the center
of mass frame, and

\begin{equation}
\gamma_c=\sqrt{s}/2m_pc^2
\end{equation}

\noindent
is the Lorentz factor of the center of mass.
Finally,
\begin{equation}
\gamma_{\pi}^{\prime}
=
{{ m_{\Delta}^2 + m_{\pi}^2 -m_p^2}\over{
2 m_{\Delta} m_{\pi} }}
\end{equation}

\noindent
is the Lorentz factor of the pion in the $\Delta$--isobar
system.

At high energies the spectrum of pions can be approximated
by the simple formula proposed by Berezinsky \& Kudryavtsev (1990):

\begin{equation}
F_{\pi}(E_{\pi},E_p)=
{1\over 2}
\left[
c_1 \left(
1 - {{ E_{\pi} }\over{E_p}} \right)^{3.5}
+c_2 \exp(-18 {{E_{\pi} }\over{E_p}})
\right],
\label{Fpi_bere}
\end{equation}

\noindent
where $c_1=1.22$ and $c_2=0.92$.
Thus the injection rate of pions is given by :

\begin{equation}
Q_{\pi}(E_{\pi^{\pm}},t)= n^p_{th} c 
\int_{p_{\rm thr}} dp N_p(p,t) \beta_p {{ F_{\pi}(E_{\pi},E_p) 
\sigma^{\pm}(p_p)}\over
{\sqrt{1 + (m_pc/p_p)^2} }},
\label{q_pi}
\end{equation}

\noindent
and the injection rate of relativistic 
electrons/positrons is given by :

\begin{equation}
Q_{e^{\pm}}(p,t)=
\int_{E_{\pi}}
Q_{\pi}(E_{\pi^{\pm}},t) dE_{\pi}
\int dE_{\mu} F_{e^{\pm}}(E_{\pi},E_{\mu},E_e)
F_{\mu}(E_{\mu},E_{\pi}),
\label{qepm1}
\end{equation}

\noindent
where, following Moskalenko \& Strong (1998), in the calculation
of the pion injection rate we combine the isobaric
model (Eqs.\ref{Fpi_steck}--\ref{f_steck}) with the
scaling model (Eq. \ref{Fpi_bere}) and adopt a linear interpolation
between the two regimes, in the energy range 3-7 GeV.
In our calculations we adopt the fits to the inclusive
cross section $\sigma^{\pm}(E_p)$ given in Dermer (1986b) which allow 
to describe separately the rates of generation of $\pi^+$ and $\pi^-$.

The pion decay is well known to generate a muon spectrum in the 
following form:

\begin{equation}
F_{\mu}(E_{\mu},E_{\pi})= 
{{ m_{\pi}^2 }\over{m_{\pi}^2- m_{\mu}^2}}
{1\over {p_{\pi}}}.
\label{fmupi}
\end{equation}

\noindent
Muons are produced in a relatively narrow range of energies, between
a kinematic mimimum and maximum given by

\begin{equation}
E_{\mu,{\rm min}}= {{ E_{\pi} }\over{ 2 m_{\pi}^2}}
\left(
m_{\pi}^2(1- \beta_{\pi}) +
m_{\mu}^2(1+ \beta_{\pi})
\right),
\label{emumin}
\end{equation}

and 

\begin{equation}
E_{\mu,{\rm max}}= {{ E_{\pi} }\over{ 2 m_{\pi}^2}}
\left(
m_{\pi}^2(1+ \beta_{\pi}) +
m_{\mu}^2(1- \beta_{\pi})
\right).
\label{emumax}
\end{equation}

\noindent
In order to speed up the computation, we assume that the spectrum
of muons is a delta--function at the energy
$E_{\mu}=1/2(E_{\mu,{\rm min}} + E_{\mu,{\rm max}})$. 
Therefore:
\begin{equation}
F_{\mu}(E_{\mu},E_{\pi})=
\delta \left(
E_{\mu} -
E_{\pi} {{ m_{\pi}^2 + m_{\mu}^2}\over{ 2 m_{\pi}^2 }}
\right).
\label{fmupi_delta}
\end{equation}

\noindent
The spectrum of electrons and positrons from the muon
decay, $F_{e^{\pm}}(E_{\pi},E_{\mu},E_e)$, was calculated by
Blasi \& Colafrancesco (1999). Combining their results with Eqs.
(\ref{qepm1}) and (\ref{fmupi_delta}), we obtain the rate of production
of secondary electrons/positrons :

\begin{equation}
Q_{e^{\pm}}(p,t) =
{{ 8 m_{\pi}^2 c}\over{ m_{\pi}^2 + m_{\mu}^2 }}
\int_{E_{\rm min}(E_e)} 
dE_{\pi} {{ Q_{\pi^{\pm}}(E_{\pi},t)}\over{E_{\pi} \beta_{\pi}}}
F_e(E_e,E_{\pi}),
\label{qepm2}
\end{equation}

\noindent
where $E_{\rm min}=2 E_e m_{\pi}^2/(m_{\pi}^2 + m_{\mu}^2)$, 
and 

$$F_e(E_e,E_{\pi})=$$
$$= {{5}\over{12}}-{{3}\over{4}}\lambda_{\pi}^2
+{1\over 3}\lambda_{\pi}^3- {{P_{\pi}}\over{2 \beta_{\pi}}}
\big( {1\over 6} - (\beta_{\pi} +{1\over 2})\lambda_{\pi}^2
+(\beta_{\pi} + {1\over 3})\lambda_{\pi}^3 \big),$$
$$\,\,\,\,\,\,\,\,\,\,\,\,\,\,\,\,\,\,\,\,
\,\,\,\,\,\,\,\,\,\,\,
\,\,\,\,\,\,\,\,\,\,\,\,\,\,\,\,\,\,\,\,
\,\,\,\,\,\,\,\,\,\,\,\,\,\,\,\,\,\,\,\,\,\,
{\rm for} \,\,\,\,\,\,\,\,
{{\gamma_{\pi}}\over{c}} (1+ \beta_{\pi})^2 > 
{{ m_{\pi}^2 + m_{\mu}^2 }\over{2 m_{\pi} E_e }};$$
and
$$= {{ \lambda_{\pi}^2 \beta_{\pi} }\over{
(1 - \beta_{\pi})^2}}\Big[
3 -{2\over 3}\lambda_{\pi}
\left({{3 + \beta_{\pi}^2 }\over{
1-\beta_{\pi} }} \right) \Big] -
{{ P_{\pi} }\over{1 - \beta_{\pi} }}
\Big\{
\lambda_{\pi}^2 (1+ \beta_{\pi}) - $$
$$\,\,\,\,\,\,\,\,\,\,\,\,\,\,\,\,\,\,\,\,
\,\,\,\,\,\,\,\,\,\,\,\,\,\,\,\,\,\,\,\,\,\,\,\,
{{ 2 \lambda_{\pi}^2}\over{1 - \beta_{\pi}}}
\left[ {1\over 2} + \lambda_{\pi}
(1 + \beta_{\pi} ) \right]
+ {{ 2 \lambda_{\pi}^3 (3 + \beta_{\pi}^2) }\over
{ 3 (1- \beta_{\pi})^2 }}
\Big\},$$
$$\,\,\,\,\,\,\,\,\,\,\,\,\,\,\,\,\,\,\,\,
\,\,\,\,\,\,\,\,\,\,\,
\,\,\,\,\,\,\,\,\,\,\,\,\,\,\,\,\,\,\,\,
\,\,\,\,\,\,\,\,\,\,\,\,\,\,\,\,\,\,\,\,\,\,
{\rm for} \,\,\,\,\,\,\,\,
{{\gamma_{\pi}}\over{c}} (1+ \beta_{\pi})^2 \leq
{{ m_{\pi}^2 + m_{\mu}^2 }\over{2 m_{\pi} E_e }}.$$

\noindent
Here $\lambda_{\pi}= 2 m_{\pi}^2 E_e/(m_{\pi}^2 + 
m_{\mu}^2) E_{\pi}$, and we put 

\begin{equation}
P_{\pi}=
- {1\over{\beta_{\pi}}}
{{ m_{\pi}^4}\over{m_{\pi}^4 - m_{\mu}^4 }}
\Big\{
4 - \big[ 1 +
\left(
{{ m_{\mu} }\over{ m_{\pi}}}
\right)^2 \big]^2
\Big\}.
\label{ppi}
\end{equation}

\noindent
Eqs.(\ref{qepm2}--\ref{ppi}) are then combined with Eq. \ref{elettroni}
to calculate the time evolution of the spectrum of the accelerated
secondary leptons.

Finally, it is useful to derive the injection rate of secondary 
electrons/positrons 
at high energies (in the scaling approximation) and for a simple
power law spectrum of cosmic ray protons, $N(E_p) = K_p E_p^{-s}$.
In this case the inclusive cross section for the production of
$\pi^\pm$ is approximately that of $\pi^o$ and thus from 
Eqs.(\ref{qepm2}--\ref{ppi}), one finds :

\begin{equation}
Q_{e^{\pm}}(p)= A(s) n_{\rm th} K_p E_e^{-s} 
\Big\{ 
{\cal P}_1 + {\cal P}_2 \ln \left({{a E_e}\over{ 6.4 {\rm GeV}}}\right)+
{\cal P}_3 \left( {{a E_e}\over{{\rm GeV}}}\right)^{1\over 2}
\Big\}
\label{qepm_s}
\end{equation}

\noindent
where $A(s)=64\times c \times 10^{-27} a^{1-s}$, and we put

\begin{equation}
a = {{ 2 m_{\pi}^2 }\over{
m_{\pi}^2 + m_{\mu}^2 }} 
\end{equation}

\noindent
and 

\begin{equation}
{\cal P}_1 = \left(
{{\tilde{a}}\over{s^2}} +
{{\tilde{b}}\over{(s+2)^2}} +
{{\tilde{c}}\over{(s+3)^2}} \right){\cal I}_o-
\left(
{{\tilde{a}}\over{s}} +
{{\tilde{b}}\over{s+2}} +
{{\tilde{c}}\over{s+3}} \right){\cal I}_1 
\label{p1}
\end{equation}

\begin{equation}
{\cal P}_2 = \left( {{\tilde{a}}\over{s}} +
{{\tilde{b}}\over{s+2}} +
{{\tilde{c}}\over{s+3}} \right) {\cal I}_o 
\label{p2}
\end{equation}

\begin{equation}
{\cal P}_3 = 1.5 \times \left( {{\tilde{a}}\over{s+1/2}} +
{{\tilde{b}}\over{s+5/2}} +
{{\tilde{c}}\over{s+7/2}} \right) {\cal I}_2 
\label{p3}
\end{equation}

\noindent
where

\begin{equation}
\tilde{a} = {5\over{12}} 
\left(
1 + {1\over 5} (a^2-1)
{{ 1 + (m_{\mu}/m_{\pi})^2}\over{
1 - (m_{\mu}/m_{\pi})^2}} \right)
\label{atilde}
\end{equation}

\begin{equation}
\tilde{b} = - {{3}\over{4}}
\left(
1 + (a^2 -1) {{ 1 + (m_{\mu}/m_{\pi})^2}\over{
1 - (m_{\mu}/m_{\pi})^2}}
\right)
\label{btilde}
\end{equation}

\begin{equation}
\tilde{c} = {{1}\over{3}}
\left(
1 + 2 (a^2 -1) {{ 1 + (m_{\mu}/m_{\pi})^2}\over{
1 - (m_{\mu}/m_{\pi})^2}}
\right)
\label{ctilde}
\end{equation}

\noindent
and where the integrals\footnote{${\cal P}_1$=0.069, 0.046,
0.031, 0.022, ${\cal P}_2$=0.019, 0.013, 0.01, 0.007,  
and ${\cal P}_3$=0.006, 0.004, 0.0033, 0.0025, 
in the case $s$=2.0, 2.1, 2.2, and 2.3, respectively} are defined as:

\begin{equation}
{\cal I}_o =
\int_0^1 {{dx }\over{x^{2-s} }}
\left[ c_1 (1-x)^{ {7\over2} } + c_2 \exp(-18 x) \right] , 
\label{i_0}
\end{equation}

\begin{equation}
{\cal I}_1 =
\int_0^1 {{dx \ln(x) }\over{x^{2-s} }} 
\left[ c_1 (1-x)^{{7\over2} } + c_2 \exp(-18 x) \right],
\label{i_1}
\end{equation}

\noindent and

\begin{equation}
{\cal I}_2 = 
\int_0^1 {{dx }\over{x^{3/2-s} }}
\left[ c_1 (1-x)^{{7\over2} } + c_2 \exp(-18 x) \right],
\label{i_2}
\end{equation}

\noindent
It follows that the slope of the spectum of secondary electrons and
positrons is essentially
that of the cosmic ray protons in this approximation (e.g., Dermer 1986a; Blasi \&
Colafrancesco 1999). In addition, Eq. \ref{qepm_s} describes the
slight departure from the simple power law shape which is due to the 
increase of the inclusive cross section with the energy of the
scattering protons (e.g., Dermer 1986b).

As shown in Paper I, the spectra of protons as affected by the
reacceleration are usually not power laws. Therefore the expressions
given here for the case of power law spectra are in general 
not applicable, although in the following we will sometimes 
use them, where specified, in order to estimate orders of magnitude.

\section{Numerical calculation of the spectra of particles and waves}

\subsection{Basic Assumptions}

A detailed modelling of the injection of MHD turbulence in galaxy
clusters and of all the related processes of wave--particle coupling is
a very complex matter and well above the capabilities of present
numerical simulations and semi--analytical treatments.
On the other hand, the basics of this process can be hopefully
understood by making use of viable assumptions and simplifications:
this is the aim of Sect.4--5.

We assume that the injection of turbulence starts in coincidence with 
a merger event and remains constant for the duration of such an 
event. As a necessary simplification the spectrum of fluid
turbulence is taken in the form of a power law (Eq.\ref{wfluid}),
which basically means that there is roughly a 
single driving scale.
The turbulence that is injected is only fluid turbulence, while
the MHD turbulence is developed later as a consequence of the 
{\it Lighthill} mechanism. We assume that the physical conditions in 
the ICM (namely magnetic field strength, temperature and number density 
of the thermal particles) do not change significantly during the time 
in which the turbulence is injected.

We also assume that there is no spatial diffusion of the particles
during the period of injection of the fluid turbulence, 
and ignore the effect of the mixing processes which may take place
during cluster merger events.

In addition we assume that the fluid turbulence and the MHD
turbulence are isotropic and that the magnetic field is 
tangled enough to ensure that also the distribution of the
accelerated particles is isotropic in pitch angle.

With these assumptions the interaction between waves and particles 
can be investigated by solving the set of coupled differential equations 
Eqs.~\ref{elettroni}, \ref{protoni}, and \ref{turbulence}. 
We consider situations in which the amount of energy injected 
in the form of turbulence is typically much smaller than the thermal energy 
of the ICM, and thus the thermal distributions of electrons and protons 
are treated as stationary. Since the time scale of damping and cascading 
are much shorter than the particle acceleration time scale, following 
Paper I we adopt a {\it quasi stationary approach}, in which it is assumed 
that within each time-step the spectrum of the waves approaches a 
stationary solution (obtained by solving Eq. \ref{turbulence} with 
$\partial W/ \partial t =0$) and that this solution changes with time 
due to the evolution of the spectrum of the accelerated electrons and 
protons.

\subsection{The wave-proton Boiler}

As discussed in Paper I, Alfv\'en waves channel most of the energy
into relativistic protons, therefore subtracting it from the electron 
component which is on the other hand responsible for the observed 
radiations. It follows that if protons are too abundant in the ICM, 
the MHD turbulence is too efficiently damped and the acceleration of electrons
is suppressed. This process is in fact quite complex, since it depends 
on the non linear time-dependent interplay among the spectra of protons, 
electrons and waves.

The aim of this Section is uniquely to draw a few general conclusions 
on the efficiency of electron and positron
acceleration in the ICM. To achieve
this goal, we simply assume that the injection rate of Alfv\'en waves
can be written as a power law, $I(k)= I_o k^{-\omega}$, and that, 
for simplicity, the spectrum of relativistic protons can also be 
approximated as a power law, $N_p(p)\propto p^{-s}$. Neither one of
these assumptions is adopted in the detailed calculations that follow.

Within the context imposed by these simple assumptions, the efficiency
of lepton acceleration can be estimated analytically.
In particular, if $\tau_s$ is the time for spectral energy transfer
due to wave-cascading and $\tau_d$ is the damping time of waves on the 
relativistic particles, one has two relevant asymptotic cases:

\begin{itemize}
\item[{\it i)}] 
For $\tau_{s} << \tau_d$, the spectrum of the waves is driven by 
the injection of Alfv\'en waves and the process of wave-cascading. 
For roughly stationary regime, the spectrum of the waves can 
be estimated from Eqs.(\ref{turbulence} and \ref{dkk}), and we can 
write:
\begin{equation}
W_k = \left( n_{\rm th} m_{\rm p} \right)^{1\over 3}
\left( {3 \over 5} {{I_{k_o} \times k_o}\over{\omega -1}} 
\right)^{2/3} k^{-{5 \over 3}}.
\label{wk_1}
\end{equation}

The lepton diffusion coefficient in momentum space is obtained 
by combining Eq. \ref{dpp} and Eq. \ref{wk_1}:

\begin{equation}
D_{pp}
\propto
\big(
{{ I_o }\over{ n_{th}}}
\big)^{ {2\over 3} }
B^{1/3}.
\label{dpp1}
\end{equation}

The efficiency of electron and positron
acceleration increases when $I_o/n_{\rm th}$ 
increases and it slightly increases with increasing $B$. Thus, if the 
turbulence is smoothly injected in the cluster volume with an injection 
rate which scales with the thermal energy density, the efficiency of 
lepton acceleration is expected to be slightly higher in the high 
field regions. On the other hand, if the injection rate of the turbulence
does not depend on the local energy density of the thermal ICM, then 
electrons are accelerated with higher efficiency in the low density 
regions.

\item[{\it ii)}] 
For $\tau_{s} >> \tau_d$, the spectrum of the waves is determined 
by the injection of Alfv\'en waves and their damping, which is mainly 
caused by resonant scattering with relativistic protons. From Eq. 
\ref{turbulence} and from the damping rate due to relativistic protons 
(Eq. 58 in Paper I) one has:

\begin{equation}
W_k= {{ c}\over{2 \pi^2 v_A^2}}
\left( {{ p_{\rm low}/(m_{\rm p}c)}\over{e\, m_{\rm p}}} \right)^{2-s}
{{ B^s}\over{{\cal E}_{\rm p}}} {{I_o s}\over{s-2}} k^{1-(\omega+s)}.
\label{wk_2}
\end{equation}

The diffusion coefficient in momentum space for electrons and positrons
can therefore be
obtained by combining Eq. \ref{dpp} and Eq. \ref{wk_2}:

\begin{equation}
D_{pp}
\propto
{{ I_o }\over{ {\cal E}_p}}
B^{1 - \omega}.
\label{dpp2}
\end{equation}
\noindent
It follows that the efficiency for lepton acceleration increases
with increasing $I_o/{\cal E}_p$ and with decreasing $B$ (at least 
for $\omega > 1$). 
\end{itemize}

\begin{figure}
\resizebox{\hsize}{!}{\includegraphics{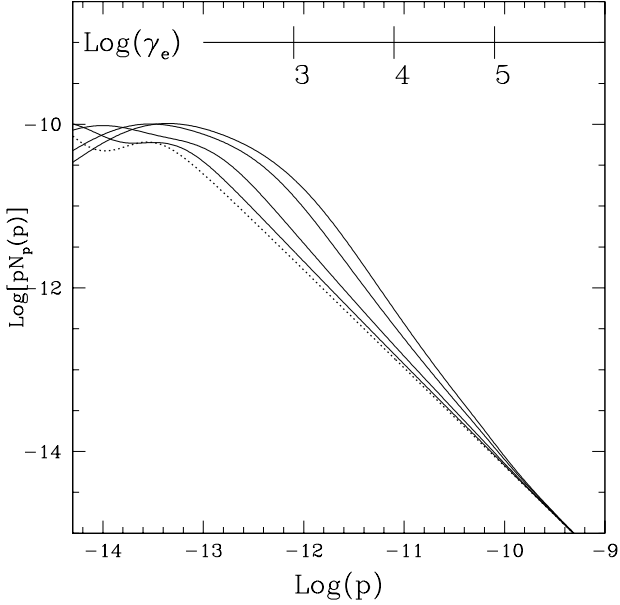}}
\caption[]{
Time evolution of the proton spectrum. From bottom to top,
we plot the spectra for $t=0$ (dotted line), $10^{15}$, $3\times
10^{15}$, $7\times 10^{15}$, and $10^{16}$ seconds after the start of 
the acceleration phase.
The calculations are carried out assuming $n_{th}=1.8\times 10^{-3}$ cm
$^{-3}$, $T=10^8$K, $B(r=0)=1.4 \mu$G, $d/dt(\delta B)^2/8\pi=1.2\times
10^{-28}$erg cm$^{-3}$ s$^{-1}$, and ${\cal E}_p=10^{-2}{\cal E}_{th}$
(with $s=2.2$).
In the Figure we also show the axis with the values of the Lorentz
factors of the secondary electrons and positrons produced by the corresponding 
cosmic ray protons.
}
\label{fig:protons}
\end{figure}

In general, the damping of waves due to resonant scattering with protons 
and the process of cascading of waves contribute rougly equally to the
modification of the spectrum of the Alfv\'en waves in the ICM, so that a
{\it realistic} situation can be thought of as intermediate between the 
two regimes discussed above, at least at 
the beginning of the reacceleration phase. 

On the other hand, when reacceleration has started, energy is transferred
from waves to protons, and the damping due to this process is expected to
increase with time. This effect is more pronounced where the Alfv\'en 
velocity is larger. As shown in Paper I, a general feature of the 
reacceleration picture in the presence of hadrons is that the 
acceleration efficiency is slightly higher in the low density
regions so that the radio 
emission due to the synchrotron radiation of electrons may have
a very
broad profile as a function of the radial distance from the cluster
center.
\begin{figure*}
\resizebox{\hsize}{!}{\includegraphics{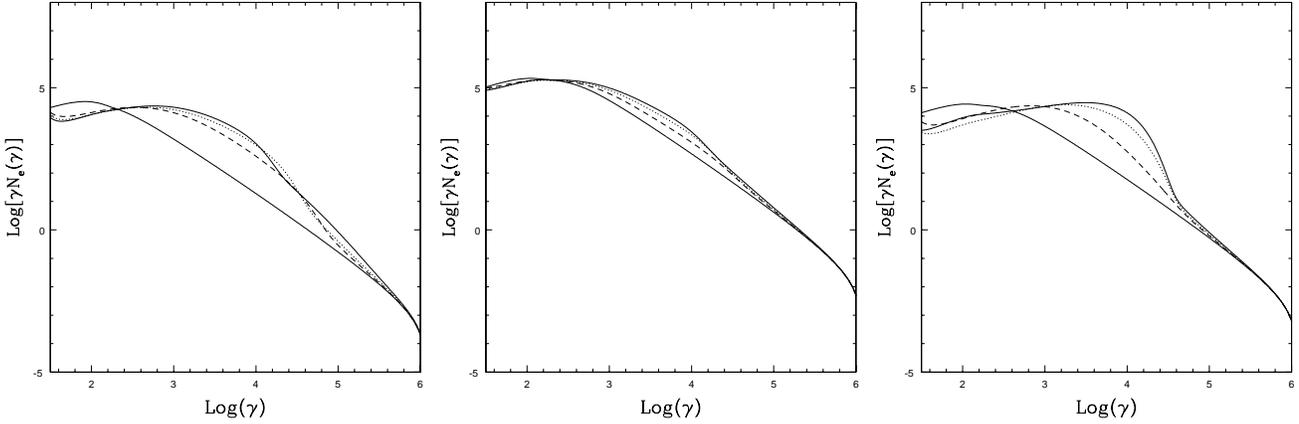}}
\caption[]{
Time evolution of the spectra of leptons.
The spectra are given at $t=0$ (thick solid lines), 
$3\times 10^{15}$ (dashed lines),
$7\times 10^{15}$ (dotted lines), 
and $1.2\times 10^{16}$ seconds (solid lines) after the start
of the acceleration phase.
{\bf Panel a)}: calculations are performed assuming
$n_{th}=2\times 10^{-4}$cm$^{-3}$ and
${\cal E}_p=8$\% of the thermal energy density.
{\bf Panel b)}: calculations are performed assuming
$n_{th}=10^{-3}$cm$^{-3}$ and
${\cal E}_p=8$\% of the thermal energy density.
{\bf Panel c)}: calculations are performed assuming
$n_{th}=10^{-3}$cm$^{-3}$ and
${\cal E}_p=1$\% of the thermal energy density.
For all the panels the other parameters are
$s=2.2$, $T=10^8$K, $B(r=0)=0.5 \mu$G and
$d/dt(\delta B)^2/8\pi=1.3\times 10^{-28}$erg cm$^{-3}$ s$^{-1}$.
}
\label{fig:3panels}
\end{figure*}

\subsection{The spectrum of relativistic protons}

Due to the confinement phenomenon and the negligible energy losses 
of hadrons in the ICM, the spectrum of the protons after the 
reacceleration phase is left basically unchanged.

In Fig. \ref{fig:protons} we illustrate an example of the time evolution 
of the spectrum of protons assuming physical conditions typical of the
core of a massive cluster. We confirm the finding of Paper I, in
which we showed that the spectrum of relativistic protons may be 
substantially modified in the energy range 1 GeV - 100 GeV
due to the resonant interaction with MHD Alfv\'enic turbulence.
This is a consequence of the fact that the acceleration time 
for protons is minimum at these energies (see Fig.~16 in Paper I).
Indeed, at smaller scales (which resonate with smaller values of
the proton energy) $\tau_s>>\tau_d$ and the acceleration time should 
decrease with particle's momentum as (from Eqs.~\ref{dpp}--\ref{kres} 
\& \ref{wk_2}):

\begin{equation}
\tau_{\rm acc}(p)
= p^3/{{\partial (p^2 D_{\rm pp})}\over{\partial p}} \propto
{{p^2}\over{D_{pp}}} 
\propto p^{-(\omega+s)+3},
\label{acctime_1}
\end{equation}

\noindent
while at larger scales (which resonate with more energetic protons)
$\tau_s << \tau_d$ and the acceleration time should increase  
with the momentum of the particles (from Eqs.~\ref{dpp}--\ref{kres} 
\& \ref{wk_1}) as:

\begin{equation}
\tau_{\rm acc}(p)
\propto p^{1/3}.
\label{acctime_2}
\end{equation}
 
In Fig. \ref{fig:protons} we also mark the typical regions of the spectrum 
of the hadrons which approximatively
contribute to the injection of the secondary electrons and positrons 
with a given Lorentz factor. The consequence of the decrease of the 
efficiency of Alfv\'en acceleration with increasing proton energy is 
that only the amount of secondary electrons/positrons injected at
$\gamma \sim 10^2-10^3$ is expected to be significantly increased.
On the other hand, the injection rate of secondary electrons and
positrons with 
$\gamma \sim 10^4$, which emit the synchrotron radiation 
at 0.3--1.4 GHz, is not expected to be substantially modified
(at least not more than a factor of 2--3) 
by the Alfv\'en acceleration process.

\subsection{Reacceleration of secondary electrons and positrons}

The main point of this paper is to include the effect of the
reacceleration of the secondary electrons and positrons, 
as generated in 
hadronic interactions of a time-dependent spectrum of protons. 
This phenomenon has a twofold effect on the observable non thermal
radiation from a cluster: first, the secondary electrons and
positrons add to
the pool of (primary) electrons that can suffer the re-energization
due to coupling with waves. Second, energy is channelled from waves 
to protons, therefore causing an increase with time of the relative 
weight of secondary electrons and positrons with respect to relic electrons.

The evolution of the spectra of particles (protons and electrons)
and Alfv\'en waves is obtained by solving numerically
Eqs.~\ref{elettroni}, \ref{protoni}, and \ref{turbulence}
with $Q_e(p_e,t)$ given by Eq. \ref{qepm2}.
The spectrum of the secondary electrons and positrons 
at the beginning of the acceleration period is computed
from the Fokker--Plank equation (Eq. \ref{elettroni},
with the source term given by Eq. \ref{qepm2})
under stationary conditions and assuming $D_{pp}=0$
(e.g., Dolag \& Ensslin 2000):

\begin{equation}
N_e(p)=
{1 \over
{\Big|
\left( {{dp}\over{dt}} \right)_{\rm rad} +
\left( {{dp}\over{dt}} \right)_{\rm i}
\Big| }}
\int_{p}^{p_{\rm max}}
Q_e(p) dp.
\label{sec_stat}
\end{equation}

Our general findings, illustrated in Fig. \ref{fig:3panels}, are 
summarized below.

As in the case of the reacceleration of relic primary electrons 
(Paper I), the efficiency for lepton acceleration decreases
with increasing energy of relativistic hadrons in the ICM.
As a consequence, the prominence of the bump of accelerated 
particles that appears in Fig. \ref{fig:3panels} is expected to 
decrease when the energy content in the form of hadrons increases.

A pronounced feature appears in the spectrum of leptons, due
to the reacceleration process, namely a sharp drop in the 
spectrum, followed by a flattening. The drop can be easily 
understood in terms of balance between energy losses of 
relativistic leptons and rate of re-energization due to
resonant interaction with waves. The flattening is simply 
due to the secondary electrons and positrons continuously generated in 
the hadronic interactions of cosmic rays in the ICM. 
Once the reacceleration period becomes longer than the 
acceleration time--scale, 
the typical energy at which this feature appears tends to
decrease with time, as a consequence of the enhanced damping of 
the waves on the proton component.
As a general comment, we point out that the presence of
reacceleration boosts the number of leptons with $\gamma \sim 10^4$
by 1--2 orders of magnitude with respect to the case 
without reacceleration.

After the end of the reacceleration stage, the spectrum of protons
remains basically unchanged. As a consequence, the spectrum of 
secondary electrons and positrons generated after the
reacceleration is stopped is
also time-independent, and determined only by the duration and by the
efficiency of the reacceleration phase.

\begin{figure}
\resizebox{\hsize}{!}{\includegraphics{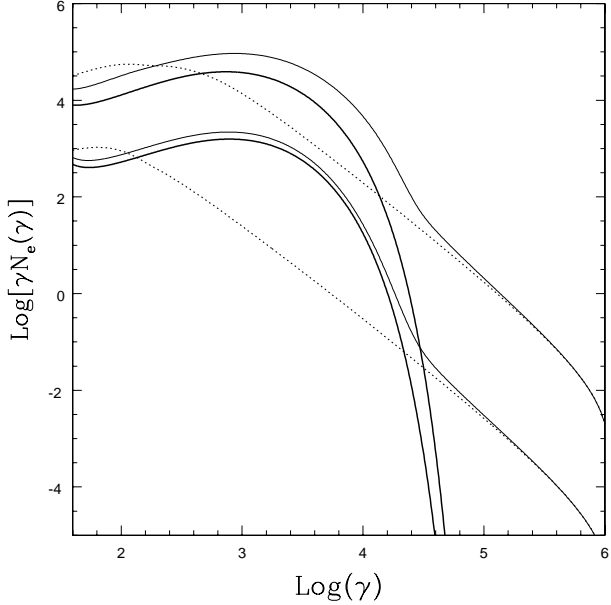}}
\caption[]{
Spectra of accelerated electrons and positrons
in a cluster core
(upper curves) and in the cluster outskirts (lower curves)
at $t=7\times 10^{15}$ s from the beginning of the acceleration
phase. Calculations in the cluster core are performed assuming 
$n_{th}=2\times 10^{-3}$cm$^{-3}$, $B(r=0)=1.5 \mu$G, and 
$d/dt(\delta B)^2/8\pi = 1.19 \times 10^{-28}$erg s$^{-1}$
cm$^{-3}$. In the cluster outskirts we assume 
$n_{th}=7\times 10^{-5}$cm$^{-3}$, $B(r=0)=0.16 \mu$G, and
$d/dt(\delta B)^2/8\pi = 7.8 \times 10^{-30}$erg s$^{-1}$
cm$^{-3}$.
${\cal E}_p=1$\% of the thermal energy density (with $s=2.2$), 
$T=10^8$K are
assumed in both the cluster core and outskirts.
The thin solid curves are the total spectra derived from the 
reacceleration of a population of relic primary electrons with 
${\cal E}_e=10^{-5} {\cal E}_{th}$ mixed with the population of 
secondary electrons and positrons.
The thick solid curves mark the contribution of the accelerated
relic electrons only.
For comparison, the stationary spectra of the secondary electrons 
and positrons in the ICM at the beginning
of the acceleration are also given in both cases (dotted lines).
The high energy cut--off in the spectrum of secondary particles is
due to a high energy cut--off fixed at 10 TeV in the spectrum of
protons.}
\label{fig:spatial}
\end{figure}

\subsection{Reacceleration of Relic and secondary electrons: {\it Hybrid 
Models}}

In Paper I we discussed at length the possibility of reaccelerating 
relic electrons, possibly injected within the cluster volume
either at shocks, or in AGNs or ordinary galaxies. When these 
injection processes take place at redshift $z<0.5$, electrons 
with Lorentz factor $\gamma \sim 200-500$ may have a lifetime of 
$\sim 10^9$ to $10^{10}$ yrs depending on whether they are injected 
in the center or in the outskirts of the cluster, respectively 
(Sarazin 1999; Brunetti 2003). It follows that, at least in the 
external parts of a cluster there may be a sufficiently high 
abundance of relic electrons to be reaccelerated by the MHD
turbulence. 

In this Section we describe the result of the reacceleration of
both components, namely relic electrons and secondary electrons
and positrons. 
We name these scenarios {\it Hybrid Models}. 

As in the previous Section 
the evolution of the spectra of particles (protons, electrons and
positrons) 
and Alfv\'en waves is obtained by solving numerically 
Eqs.~\ref{elettroni}, \ref{protoni}, and \ref{turbulence}
with $Q_e(p_e,t)$ given by Eq. \ref{qepm2}.
In this case, however, the initial spectrum of the electrons 
is the combination of the spectrum of the secondary electrons
and positrons 
at the beginning of the acceleration period (Eq. \ref{sec_stat}) 
and of the spectrum of the relic--primary electrons accumulated
in the ICM.

In Fig. \ref{fig:spatial} we plot the spectrum of the reaccelerated 
electrons and positrons 
as obtained in the cluster center and in the cluster outskirts 
(see the caption for the numerical values adopted there).
In Fig. \ref{fig:spatial} we assume that the strength of the magnetic 
field in the cluster volume scales according with flux conservation 
($B \propto n_{\rm th}^{2/3}$) and that the injection power of Alfv\'en 
waves is $P_A = \int I_k dk \propto n_{\rm th}^{5/6}$ (see Sect.~5.1 and
Paper I). The number density of the relic (primary) electrons and 
of the relativistic hadrons in the cluster volume scales with that 
of the thermal particles. Assuming that the energy density of the
relativistic protons is of the order of $10^{-2}$ times that of the
thermal plasma, from Fig. \ref{fig:spatial} it follows that
the reacceleration process of secondary electrons/positrons 
could be important.
In particular, we find that a relevant contribution (from $\sim$50 to 
80\%) to the spectrum of the radiating electrons and positrons with 
$\gamma \sim 10^3-10^4$ in the central regions of the cluster is 
provided by reaccelerated secondary electrons and positrons. 
On the other hand, 
the spectrum of electrons in the external regions is essentially 
contributed by the reaccelerated relic electrons. In the presence of
processes that make the spatial distribution of relativistic protons
broader than the thermal gas, we can expect that the contribution
provided by the central denser regions of the cluster gets somewhat
suppressed. At the same time, since in the cluster outskirts the
Coulomb losses of the relativistic electrons are less efficient than 
in the cluster center, the radial distribution of the number density 
of relic  electrons may be broader than that assumed in Fig. \ref{fig:spatial} 
and consequently their contribution to the total spectrum of the 
reaccelerated leptons may be even larger.

\section{Hybrid Models: phenomenology of non thermal emission from 
galaxy clusters}

Here we apply the formalism described in the previous Sections and 
calculate the expected non thermal emission from galaxy clusters,
when both relic electrons and secondary electrons and positrons
are
present during the stage of injection of turbulence and resonant
reacceleration.

\subsection{Basic Assumptions}

In this Section we briefly discuss the basic assumptions adopted for 
the calculations of particle acceleration and non--thermal emission
from galaxy clusters. The assumptions are relative to the physical
properties of the ICM and of the relativistic component, and to
the modelling of the injection of turbulence in the cluster volume.

We assume a $\beta$-{\it model} (Cavaliere \& Fusco-Femiano, 1976) for 
the radial density profile of the thermal gas in the ICM, in the form
\begin{equation}
n_{th}(r)=n_{th}(r=0)
\Big(
1+
( {{ r }\over{r_c}} )^2
\Big)^{-3\beta /2},
\label{betamod}
\end{equation}
where $r_c$ is the core radius and $\beta = 0.8$. The magnetic field is 
taken in its flux conserving form:
\begin{equation}
B(r) = B(r=0)
\Big(
{{ n_{th}(r) }\over{
n_{th}(r=0) }}
\Big)^{2/3}.
\label{b(r)}
\end{equation}
Here we explore the region of values $B(r=0)\sim$0.5--3$\mu$G, expected to
reproduce the uncertainty in the value of the magnetic field as derived
from different techniques.

Following Fujita et al.(2003) and paper I, we assume that large scale fluid 
turbulence is injected in the ICM during cluster mergers and that the 
turbulent eddies at small scales radiate MHD waves due to the {\it Lighthill} 
mechanism (Sect. 3.2.3). 

For simplicity we assume that the maximum injection scale of the turbulence, 
the Reynolds number and the velocity of the turbulent eddies, which are 
essentially unknown quantities, are independent from the location within the 
cluster volume. Under these simplified conditions, in Paper I we showed that 
the injection power in the form of Alfv\'en waves scales as:
\begin{equation}
P_A(r)=
\int I_k(r) dk =
P_A(r=0)
\Big(
{{ n_{th}(r) }\over{
n_{th} (r=0) }}
\Big)^{5/6}.
\label{pa(r)}
\end{equation}
Finally we assume that the spatial profile of the number density
of the relic electrons and of the relativistic protons (at the 
beginning of the acceleration period) scales with that of the 
thermal matter:
\begin{equation}
{\cal E}_{p[e]}
={\cal E}_{th}
\eta_{p[e]},
\label{ratioenerg}
\end{equation}
where $\eta_{p[e]}$ is a parameter; a reference value, $s=2.2$,
is adopted for the injected spectrum of the cosmic ray protons.
As a matter of fact the use of Eq.(\ref{ratioenerg}) 
limits ourselves to assume a fairly uniform distribution 
(non--patchy) of relativistic particles in the ICM.
On the other hand, patches of relativistic plasma separated from
the thermal pool are
clearly detected by radio and X--ray observations (e.g.,
Fabian et al. 2000; McNamara et al., 2000) 
which show the presence of bubbles and cavities in a few galaxy 
clusters.
However, it is expected that such bubbles will expand with time
and mix in the ICM due to the developing of instabilities 
in a time scale of the order of a few
$10^8$yrs (Churazov et al., 2000; Br\"uggen \& Kaiser 2001) which
is a short time scale with respect to the duration of the
injection process of the bulk of cosmic rays in galaxy clusters.
Thus the assumption of a fairly uniform mixing between thermal
and relativistic plasma in galaxy clusters is justified for
the aim of the present paper.

\subsection{Particle acceleration in cluster cores}

In this Section we focus on the synchrotron emission expected from 
the cores of galaxy clusters and derive constraints on the physical 
properties of the ICM. We focus on cluster cores since 
in these regions the larger gas density makes
the $pp$ interactions more frequent and therefore the density of
secondary particles larger. Moreover the magnetic field is expected
to be larger in the center. These two facts imply a larger synchrotron
emissivity in these regions. 
On the other hand, if relativistic protons 
have more than a few percent of the local thermal energy density, the
damping of waves becomes too large and the electron acceleration gets 
suppressed, therefore reducing the synchrotron emissivity. It follows 
that the general situation may be rather complex.

\begin{figure*}
\resizebox{\hsize}{!}{\includegraphics{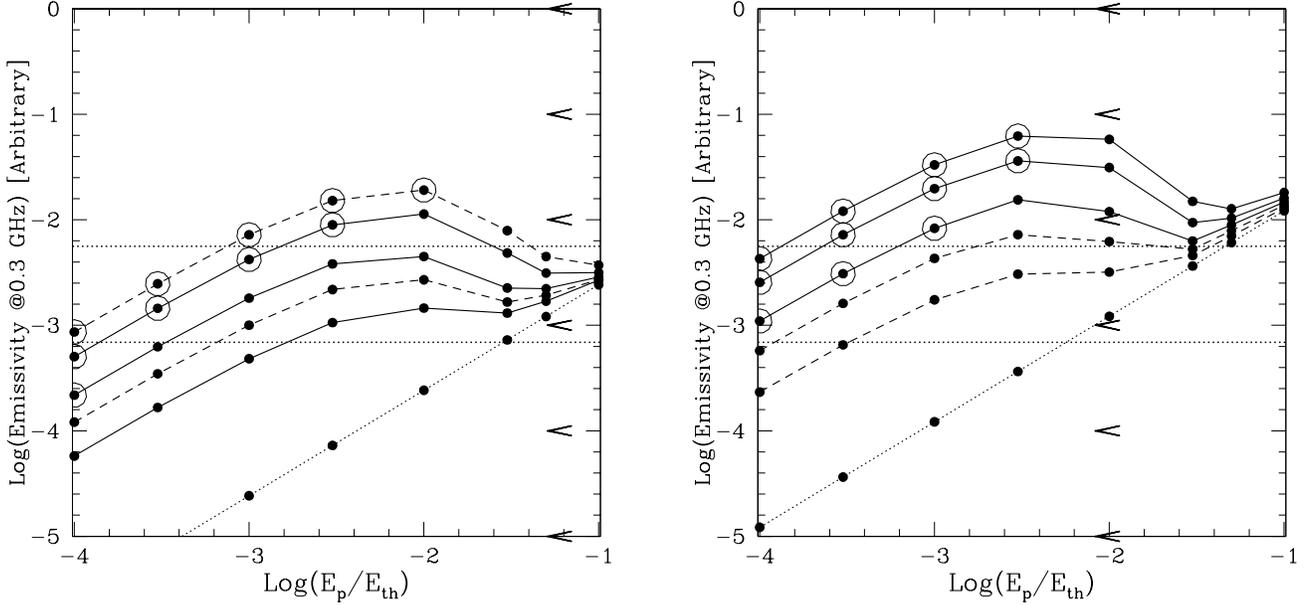}}
\caption[]{
Synchrotron emissivity at 330 MHz (in arbitrary units) as
a function of the ratio between cosmic rays and thermal
energy densities at the beginning of the reacceleration stage
(${\cal E}_p/{\cal E}_{th}$) 
for the core region of the Coma cluster calculated after a
reacceleration period $\Delta \tau_{acc}=10^{16}$ s.
The horizontal dotted lines give the upper and lower bounds 
of the allowed range of the synchrotron
emissivity obtained from the observed brightness profile at
330 MHz (estimated from the deprojection of the
observed profile given by Govoni et al. 2001).
Arrows gives the limits on the ratio ${\cal E}_p/{\cal E}_{th}$ 
obtained from the high frequency radio data (Reimer et al. 2004) 
and assuming that all the emission at these frequencies comes from the
core region.
Circles mark the points in the diagram for which the synchrotron
spectral index is consistent with that observed 
(from Giovannini et al.~1993). 
{\bf Panel a)}:
Calculations are carried out for $B(r=0)=1\mu$G.
The different curves refer to different assumed injection rates
for Alfv\'en waves ($d/dt(\delta B)^2/8 \pi$):
1.87, 1.3, 0.83, 0.64, and $0.47\times 10^{-28}$erg s$^{-1}$
cm$^{-3}$ (from top to bottom, respectively).
{\bf Panel b)}:
Calculations are carried out for $B(r=0)=3\mu$G.
The different curves refer to different assumed injection rates
for Alfv\'en waves ($d/dt(\delta B)^2/8 \pi$):
1.3, 0.83, 0.47, 0.33 and $0.21\times 10^{-28}$erg s$^{-1}$
cm$^{-3}$ (from top to bottom, respectively).
}
\label{fig:2panels}
\end{figure*}

This complexity is illustrated in Fig. \ref{fig:2panels} where we plot 
the synchrotron emissivity as a function of the ratio of the energy 
densities in the relativistic protons and the thermal plasma
at the beginning of the reacceleration stage, 
${\cal E}_p/{\cal E}_{th}$, 
for different values of $B$ and of $P_A$.

As expected, for small values of ${\cal E}_p/{\cal E}_{th}$ 
the damping rate
is less efficient than the cascading process and the synchrotron 
emissivity simply scales with ${\cal E}_p/{\cal E}_{th}$ 
(Eq. \ref{dpp2}); at this stage, for the adopted rates of
turbulence--injection, the synchrotron emission is more than one
order of magnitude larger than in the case without reacceleration.
Increasing ${\cal E}_p/{\cal E}_{th}$, 
the damping of waves on protons increases and the acceleration efficency 
decreases, so that the synchrotron emissivity is reduced as well.
Fig. \ref{fig:spatial} also shows that when the ratio 
${\cal E}_p/{\cal E}_{th}$ is larger 
than $\sim 10-20\%$, then the reacceleration of leptons with $\gamma 
\sim 10^4$ is basically stopped and the synchrotron emissivity approaches 
that expected from the standard secondary model.

When this saturation effect does not occur, the synchrotron emissivity
in the central region may easily exceed the observations. As a result, 
such observations can be used to impose contraints on the physical 
conditions in which the reacceleration of secondary particles takes
place. As usual we refer to the case of the Coma cluster as the case
in which the observations are richer. The conclusions that we will 
draw below should not be extended to other clusters, until 
comparable wealth of data is obtained for those clusters.

The main observational constraints can be summarized as follows:

\begin{itemize}
\item[{\it i)}] {\it Radio brightness of the core region}:
we use the 327 MHz VLA profile obtained after the subtraction of
point-like sources (Govoni et al.~2001). The allowed region of the
brightness of the core 
region can be estimated by subtracting the contribution due to the 
external regions to the brightness integrated along the line of sight.

\item[{\it ii)}] {\it Radio spectrum of the core region}:
the 327-1400 MHz spectral index map of the Coma radio halo shows
a plateau in the central regions and a prominent radial steepening 
of the slope of the spectrum (Giovannini et al. 1993). The plateau 
region roughly coincides with the cluster core and the sychrotron 
spectrum there is slightly flatter than $\alpha \sim 0.9$.

\item[{\it iii)}] {\it Radio brightness in the frequency range 2.7-5 GHz}:
At these high frequencies the radio emission imposes strong constraints on
all flavors of secondary models, as discussed by Reimer et al.(2004).

\end{itemize}

In Fig. \ref{fig:2panels} we compare our theoretical expectations, as 
obtained for different values of the model parameters ($P_A$, $B$ and 
${\cal E}_p/{\cal E}_{th}$), with the constraints listed above. 

The weakest limits are clearly those obtained for low values of the 
magnetic field. For $B=1\mu G$ (left panel in Fig. \ref{fig:2panels}), 
low rates of injection of waves select the region with 
values of ${\cal E}_p/{\cal E}_{th}$ larger than about $10^{-3}$.
On the other hand, as soon as the rate of injection increases above 
$\sim 8 \times 10^{-29}$erg cm$^{-3}$ s$^{-1}$ (which implies a total
energy budget injected per unit volume in the form of 
Alfv\'en waves $\sim 1.5$\% of the 
thermal energy density), the allowed values of 
${\cal E}_p/{\cal E}_{th}$ drop below a few $10^{-3}$. 
The high frequency data of the Coma cluster 
impose an upper limit ${\cal E}_p/{\cal E}_{th}<0.05$. 

Assuming $B=3\mu G$ (right panel in Fig. \ref{fig:2panels}) it is clear 
that the observed 
synchrotron brightness excludes big chunks of the parameter space. In 
particular, assuming an appreciable injection rate of energy of Alfv\'en 
waves in the cluster core ($\geq 0.5$\% of the thermal energy) it is found 
that intermediate values of the ratio ${\cal E}_p/{\cal E}_{th}$ 
are not allowed.
The region ${\cal E}_p/{\cal E}_{th} > 10^{-2}$
is excluded by the high frequency
points in the spectrum of the radio halo of the Coma cluster (arrows in
Fig. \ref{fig:2panels}).

The most difficult observational constraint to match is the combination
of low synchrotron brightness {\it i)} and flat radio spectrum {\it ii)}. 
A relatively low value of the acceleration efficiency 
cannot reproduce a 
synchrotron spectrum as flat as the observed one. Therefore an efficient 
particle acceleration mechanism is requested to boost electrons toward 
higher energies and to flatten the emitted synchrotron spectrum. In order
to avoid to exceed the observed brightness, a relatively small injection 
rate of secondary electrons and positrons
is required. More quantitatively, we find that
the parameter
space with ${\cal E}_p/{\cal E}_{th} > 10^{-3}$ 
is excluded for this strongly magnetized
case.

\subsection{Integrated broad band spectrum}

In this Section we illustrate our calculations of the volume integrated
fluxes of radiation generated by reaccelerated electrons and positrons
through 
synchrotron emission and IC. The central gas density, $n_{th}(r=0)$, 
the $\beta$ parameter and the core radius $r_c$ are chosen as the 
representative values of the Coma cluster (Briel et al. 1992).

\begin{figure*}
\resizebox{\hsize}{!}{\includegraphics{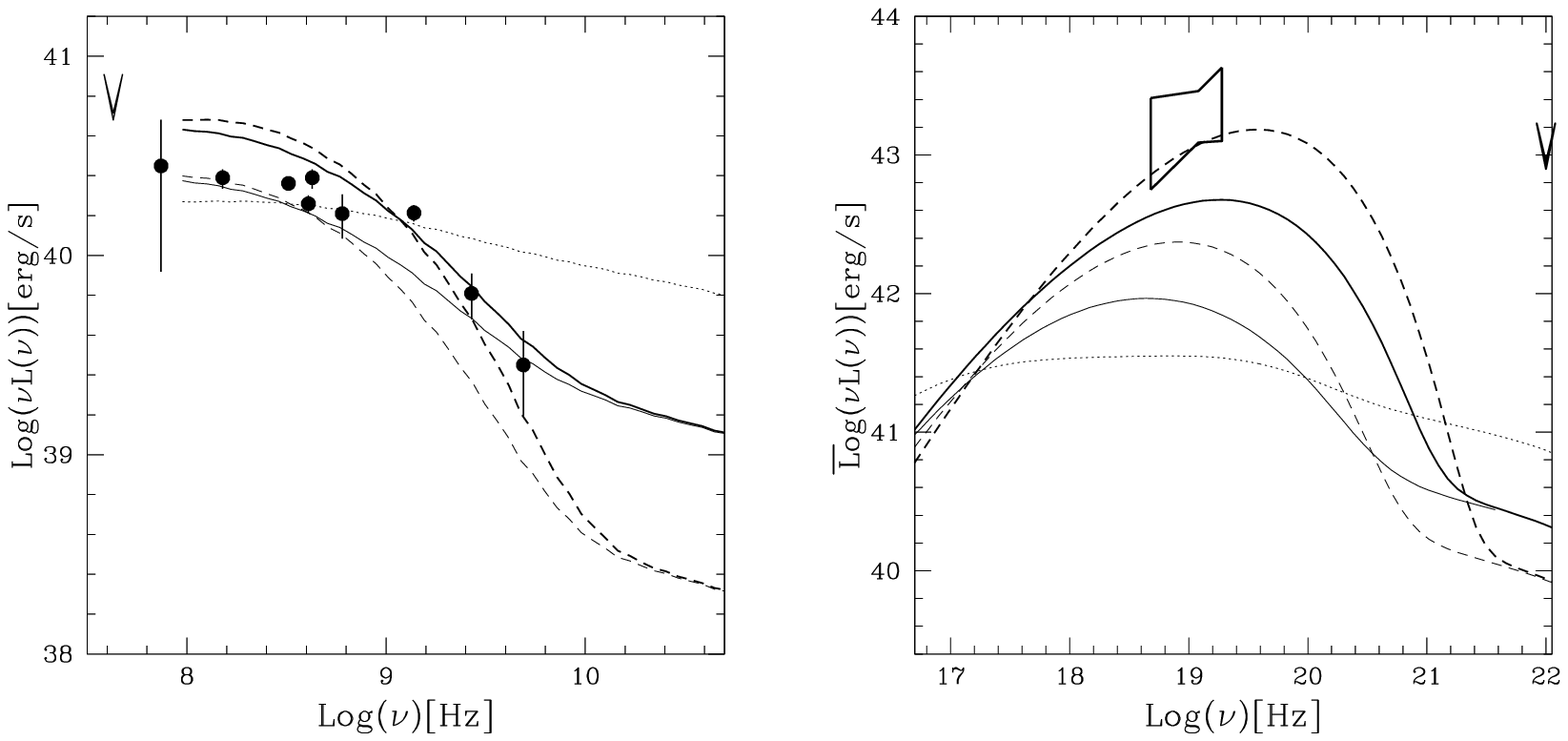}}
\caption[]{
Expected synchrotron (left panel) and IC (right panel) spectra 
calculated for the Coma cluster.
All the calculations are carried out assuming that the injection 
rate of Alfv\'en waves in the center of the cluster is
$P_A(r=0) = 1.3 \times 10^{-28}$erg s$^{-1}$ cm$^{-3}$ and for
$\Delta t_{acc}=10^{16}$ s.
Calculations are carried out for both 
$P_A \propto n_{th}^{5/6}$ (Eq. \ref{pa(r)}) (thin lines)
and $P_A \propto n_{th}^{1/2}$ (thick lines).
The different models refer to the following cases: $B(r=0)=2.0 \mu$G and
${\cal E}_p/{\cal E}_{th}=0.12$
with no relic primary electrons (dotted lines),
$B(r=0)=1.7 \mu$G and
${\cal E}_p/{\cal E}_{th}=0.03$
with no relic primary electrons (solid lines),
and 
$B(r=0)=1.1 \mu$G and
${\cal E}_p/{\cal E}_{th}=0.01$
with ${\cal E}_e/{\cal E}_{th}=5 \times 10^{-5}$ independent
of the distance from the center (dashed lines).
}
\label{fig:broad}
\end{figure*}

In Fig. \ref{fig:broad} we plot our results for the synchrotron spectra (left
panel) and the IC spectra (right panel). The data points refer 
to the radio, hard X-ray and gamma ray bands. All curves are 
obtained in the assumption that the cosmic ray energy density 
at the beginning of the reacceleration stage
is proportional to the thermal energy density at any point.
The values of $P_A(r=0)$ and the ratio 
${\cal E}_p/{\cal E}_{\rm th}$ are
not chosen to obtain a best fit to the data, they are only fixed
in order to provide a viable representation of the data.

Some general remarks emerge from the inspection of Fig. \ref{fig:broad} :
\begin{itemize}
\item[{\it i)}] The synchrotron luminosity of a Coma--like
radio halo can be easily matched even with ${\cal E}_p/{\cal E}_{\rm th}
\sim 10^{-3}-10^{-2}$ and $\sim \mu$G strengths of the central magnetic
field.
This energy requirement is more than one order of magnitude
below that of classical secondary models (Blasi \& Colafrancesco
1999; Dolag \& Ensslin 2000).

\item[{\it ii)}] The steepening of the integrated synchrotron
spectrum of the Coma radio halo can be reproduced only for cosmic 
ray energy density of the order of ${\cal E}_p \leq 5$\% 
or lower. This is because a larger content of cosmic ray protons would
decrease the efficiency of the lepton acceleration and 
reduce the synchrotron bump at lower frequencies.

\item[{\it iii)}] The observed hard X-ray spectra are hardly achievable 
if only the effect of reaccelerated secondary particles is taken 
into account. This is mainly due to the low number of secondaries 
generated if the energy density in the form of relativistic protons
is the one inferred in our point {\it i)}. 

If the injection of waves takes place on a spatial scale which 
is appreciably broader than that in Eq. \ref{pa(r)},
the efficiency of the reacceleration of secondary
particles increases in the external volume and 
the IC emission in the outskirts can be enhanced,
leaving the synchrotron emission almost unaffacted, due to the
rapid decrease of the value of the magnetic field with radius. 
However, we find that a flux of HXR close to the observed one can 
be obtained only by assuming rather extreme conditions in the cluster 
outskirts (e.g., ${\cal E}_t \sim {\cal E}_{th}$).
Furthermore we find that in this case 
the strong back--reaction 
of the accelerated protons would suppress
the acceleration of electrons within $\sim 10^8$yrs.

Similarly an appreciably larger IC luminosity cannot be produced
if the spatial distribution of the cosmic ray protons
is broader than that of the thermal plasma.
Indeed, although in this case a larger number of 
secondary electrons and positrons is produced in the external volume, 
the stronger back reaction of protons on the waves inhibits 
the acceleration of electrons/positrons in these regions.

\item[{\it iv)}] Due to the relatively poor efficiency of the 
reacceleration mechanism, Alfv\'en waves in the ICM cannot accelerate 
very high energy electrons (say $\gamma \geq 10^5$) and thus the 
energy distribution of the electrons and positrons 
which are responsible for the 
emission of gamma rays though IC is not appreciably affected by the 
reacceleration scenario discussed here. This is the reason why in the
pure reacceleration models, with no protons and no secondary particles,
we expect no gamma ray emission. The limit imposed on the energetic 
budget in the form of high energy protons from the EGRET upper limit 
(Reimer et el. 2004) is at the level of 20\% of the thermal energy, 
obtained by assuming that the all gamma ray flux is generated through 
pion decays. For cosmic ray energy densities below this bound
(but larger than a few times $10^{-3}$ of the thermal energy), an
anti-correlation appears between the IC HXRs and the IC gamma rays
(we recall that the latter are only generated by the secondary 
electrons and positrons); a similar anti-correlation is expected between
the IC HXRs and the gamma rays generated by the decay of $\pi^o$. 
A detailed analysis of the gamma ray emission expected from
reacceleration models will be presented in a forthcoming 
paper (Brunetti et al., in prep.).

\item[{\it v)}] In principle the broad band non--thermal spectrum 
of the Coma radio halo can be reproduced if both relic electrons and
secondary electrons/positrons 
are present\footnote{For simplicity we assume 
Eq. \ref{ratioenerg} to hold also in the case of relic electrons.}.
If a few percent of the thermal energy are stored in the relativistic
protons, the synchrotron spectrum may be dominated by the radiation 
from reaccelerated secondary electrons and positrons, 
mainly in the central regions
of the cluster. At the same time, the IC emission is dominated by
reaccelerated relic electrons in the external regions (i.e. at 
$2 < r/r_c < 5$).

Obviously the contribution of the reaccelerated
secondary electrons and positrons to the integrated 
non--thermal spectrum of galaxy clusters
is expected to fall down if the energy of cosmic ray
protons is maintained well below $\sim 1$\% of the thermal energy.
\end{itemize}

\subsection{Radial profiles}

The very broad extension of the synchrotron 
emission from giant radio halos is among the
properties which are difficult to be fitted 
by secondary models (e.g., Brunetti 2004 and ref.
therein).
Although this Section is not devoted to a detailed
comparison between observed synchrotron profiles and model 
expectations, here we show that Alfv\'enic
reacceleration of secondary electrons and positrons in the ICM may 
generate relatively broad synchrotron profiles.

If the reacceleration period is much longer than the 
reacceleration time--scale, the bulk of the secondary electrons
and positrons  
injected above the momentum, $p_{_{_{_>}}}$, 
at which Coulomb losses outweight the acceleration efficiency, 
is essentially boosted around a maximum
momentum, $p_{\rm max}$, 
at which acceleration is balanced by radiative losses.

From Figs.~12 \& 13 of Paper I one finds that under the
assumed physical conditions the acceleration of the lower
energy electrons and positrons 
typically happens in the regime 
$\tau_s >> \tau_d$ while the acceleration of the higher energy
leptons happens in the opposite regime. 

Thus, assuming for simplicity a power law energy distribution
of the relativistic protons $N_p = K_p p^{-s}$, from 
Eqs.(\ref{ion}), (\ref{dpp}), (\ref{wk_2}), and (\ref{acctime_1}) 
one finds:

\begin{equation}
p_{_{_{>}}}
= \left(
{{ A_{\rm C} }\over{A_{\rm w}}}
{{n_{\rm th} {\cal E}_{\rm p}}\over
{I_o}} B^{\omega -1} 
\right)^{1/(\omega + s -2)}
\label{p>}
\end{equation}

\noindent
where $A_C$ is the constant in Eq. \ref{ion}, and 

\begin{equation}
A_{\rm w}=
{{ 2(1+\omega +s)}\over{s-2}}
{{ ({\rm e}/c)^{3-(\omega +s)} }\over
{(\omega +s )^2 -1 }}
\left(
{{ p_{\rm low} /m_{\rm p} c }\over
{{\rm e} m_{\rm p}}}
\right)^{2-s}
\label{aw}
\end{equation}

\noindent
while from Eqs.(\ref{syn+ic}), (\ref{dpp}), (\ref{wk_1}),
and (\ref{acctime_1}) one has:

\begin{equation}
p_{\rm max}
= \left(
{{ A_{\rm ww} }\over{A_{\rm rad}}}
{1 \over{ B_{\rm IC+}^2}}
\right)^{3/4}
\left(
{{ I_o B^{1/2}}\over{n_{\rm th}}}
\right)^{1/2},
\label{pmax}
\end{equation}

\noindent
where $A_{\rm rad}$ is the constant in
Eq. \ref{syn+ic}, 
$B_{\rm IC+}^2 = B_{\rm IC}^2 +B^2$, and 

\begin{equation}
A_{\rm ww}=
{{3}\over{5}} \pi {\rm e}^{1/3}
\left(
{{ 3 m_{\rm p} c^2 }\over{
5(\omega -1)}}
\right)^{2/3}.
\label{aww}
\end{equation}

\noindent
The number density of the reaccelerated secondary electrons
and positrons around $p_{\rm max}$ can be estimated 
by the integral of the number density of secondary particles
injected with $p > p_{_{_{>}}}$ during the reacceleration
stage. From Eqs.(\ref{qepm_s}) (neglecting for simplicity
the contributions from the ${\cal P}_2$ and ${\cal P}_3$
terms) and (\ref{p>}), one has :

\begin{equation}
N^+_{{\rm e}^{\pm}}
\sim {{\Delta_T A(s) c^{-s} {\cal P}_1}\over{
s -1}} n_{\rm th} K_{\rm p}
\left(
{{A_{\rm C} }\over{ A_{\rm ww}}}
{{ n_{\rm th} {\cal E}_p B^{\omega -1}}\over
{I_o}}
\right)^{ {{1-s}\over{\omega +s -2}} }.
\label{neppm}
\end{equation}

\noindent
On the other hand, the number density of electrons with $p \sim p_{\rm max}$
in the classical secondary model 
can be obtained from Eqs.(\ref{qepm_s}), (\ref{sec_stat}) 
and (\ref{pmax}) :

\begin{equation}
N_{{\rm e}^{\pm}} 
\sim {{ p_{\rm max}^{-(s+1)} }\over
{s -1}} {{ A(s) c^{-s} {\cal P}_1 }\over{A_{\rm rad}}}
{{ n_{\rm th} K_{\rm p} }\over{
B_{IC+}^2 }}
\label{nepm}
\end{equation}

\noindent
Thus the increase of the extension of the emitted synchrotron
profile (associated to electrons with $p\sim p_{\rm max} {\rm c}$)
in the secondary--reacceleration model with respect to that in the 
classical secondary model can be directly estimated from the ratio:

\begin{equation}
N^+_{{\rm e}^{\pm}}/N_{{\rm e}^{\pm}}
\propto
\left( {{ n_{\rm th} {\cal E}_p B^{\omega -1}}\over
{I_o}} \right)^{ {{1-s}\over{\omega +s -2}} }
\left(
{{ I_o B^{1/2} }\over{ n_{\rm th} }}
\right)^{ {{s +1 }\over{2}} },
\label{ratio_n}
\end{equation}

\noindent
which, in the assumption that ${\cal E}_p \propto n_{\rm th}$,
roughly scales as $N^+_{{\rm e}^{\pm}}/N_{{\rm
e}^{\pm}} \propto n_{\rm th}^{-1/2}$-- $n_{\rm th}^{-3/2}$.
Eq. \ref{ratio_n} provides a simple way to estimate the increase 
of the extension of the radial profile (at least for $s\sim 2-2.3$).
However, the injection spectrum of secondary electrons/positrons 
at lower energies
(i.e., $\gamma < 10^3$) is not well described by a simple
power law (Eq. \ref{qepm_s}), and the spectra of protons as 
affected by reacceleration are not power laws either.
Thus a more detailed calculation is required.
In Fig.~6 we plot the ratio of the synchrotron emissivities as given
with reaccelerated particles and in the context of classical secondary
models (the parameters are as used in Fig.~5). It is clear that the models
invoking reacceleration of secondary particles may produce broader
synchrotron emission with respect to secondary models.
However, we also notice that the presence of a cut--off in the
synchrotron spectrum as obtained in reaccelerated models (which occurs
at lower frequencies with increasing radius) produces a steepening 
of the synchrotron spectrum which balances the increase of the ratio 
$J_{Syn}^+/J_{Syn}$ with the distance from the center.

\begin{figure}
\resizebox{\hsize}{!}{\includegraphics{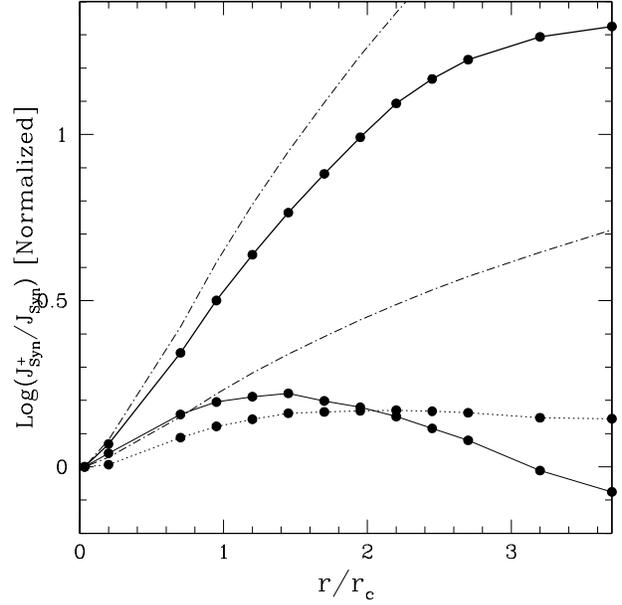}}
\caption[]{
Normalized ratio between the synchrotron emissivity
(at 330 MHz) produced by reaccelerated secondary particles,
$J_{Syn}^+$, and that produced by the stationary injection 
of secondary particles, $J_{Syn}$, as a function of the
distance from the cluster center (in units of the core radius).
The reacceleration models are the same as in Fig.~5 
(with the exception of the {\it hybrid} models shown as
dashed lines in Fig.~5) and are shown with the same symbols.
The synchrotron emissivity in the classical secondary
model is computed in the assumption that $B(r=0) << 3 \mu$G.
The dot--dashed lines reproduce the behaviour $\propto
n_{\rm th}^{-1/2}$ and $\propto n_{\rm th}^{-3/2}$
as expected from Eq. \ref{ratio_n}.
}
\label{fig:prof}
\end{figure}

In order to reproduce the extension of the largest
radio halos with the assumptions in Sect.~5.1,
the synchrotron emissivity in the external regions should be
amplified by a factor of $\sim 100$ with respect to that 
produced in classical secondary models (e.g.,
Brunetti 2004, Fig.~2); this large factor is not obtained in our
calculations (Fig.~6).
In order to obtain a very broad synchrotron radio emission,
the radial distribution of the energy density of 
cosmic ray protons can be forced to be more extended than  
that of the thermal ICM.
However, in this case the synchrotron profile is not broadened 
enough because at large distances the damping of the Alfv\'en waves 
due to cosmic rays gets stronger.

\section{Discussion and Conclusions}

We presented the results of the first self-consistent calculations
of the injection and reacceleration of cosmic ray protons and both
primary and secondary leptons in the intracluster medium. This goal
is reached by solving the coupled equations for the time-dependence
of the protons, electrons/positrons 
and Alfv\'en waves spectra. The work 
presented here is the natural continuation of a previous work
(Brunetti et al. 2004). In particular, in the present paper we
illustrate novel results concerning the role of the injection 
and further re-energization of secondary particles due to their
resonant interaction with Alfv\'en waves. Such secondary particles
must exist in the ICM, at least to some extent to be defined: the
reason for such a certainty is that we are aware that there are
cosmic ray sources in clusters and that the bulk of cosmic rays generated
by such sources are diffusively trapped within the ICM (V\"{o}lk et
al. 1996; Berezinsky, Blasi \& Ptuskin 1997). However, neither the
amount of energy in the form of cosmic rays, nor their spectrum or
their spatial distribution within the ICM are currently known. Their
production could be related to the formation of the large scale 
structure of the universe or to the astrophysical sources within 
clusters (galaxies and AGNs). 

The main assumption of our calculation is the choice of a specific  
type of MHD waves, namely Alfv\'en waves. In general one could 
think of any combination of different modes which behave differently
in their interaction with particles (electrons and protons). In 
particular, the case of magnetosonic waves, considered by Cassano
\& Brunetti (2005) appears to be particularly interesting since 
in that case the problem of wave injection on small spatial scales 
is alleviated. In the present paper, the presence of Alfv\'en waves
on small scales is solved by assuming that fluid turbulence is 
injected on large scales and gets eventually coupled with Alfv\'en
turbulence on small scales through the so-called {\it Lighthill}
mechanism. Obviously, direct proofs that this process may be at work
require MHD simulations with resolution and dynamical range which are not 
available at present. 

The wave-particle coupling implies a relevant effect on the presence
of secondary particles, mainly in two ways: first, the spectrum of 
parent cosmic rays is changed by the reacceleration process (as also
found in Paper I) therefore affecting the spectrum of secondary 
electrons and positrons. 
Second, the secondary electrons and positrons
are in turn re-energized
by the resonant interaction with Alfv\'en waves. The effect of the 
reacceleration on the confined cosmic rays is shown in Fig. 1 and
most notably consists of a bump at Lorentz factors below $\sim 10^2$,
which becomes increasingly more evident with time after the start of
the reacceleration phase. At larger Lorentz factors the spectrum of 
cosmic rays remains unchanged. A similar bump shows in the spectrum
of the total population of electrons and positrons
(primaries plus secondaries)
as plotted in Fig. 3. The pumping of energy into relativistic 
particles through the resonant interactions with Alfv\'en
waves at some point produces an interesting effect, which in Paper I 
we named {\it wave-proton boiler}: the reacceleration of electrons 
and positrons 
continues provided the wave damping on the proton component is not
too large; when the energy present in the form of protons exceeds
some limit the reacceleration of electrons gets suppressed. In
this sense the system made of protons, electrons and waves is
self-regulated. 

One of the most common criticisms to the so-called {\it
reacceleration} scenarios is that the origin of the relic
electrons to start with is left as an open issue: we think
that the presence of the continuously generated secondary
particles largely mitigates this problem, since $pp$ 
interactions continuously inject these particles in the
ICM. Moreover, astrophysical sources such as the lobes
of radio galaxies
and active galaxies are {\it seen} to pollute the ICM with 
a population of relativistic electrons, therefore it is
plausible that both leptons of primary and secondary 
origin may be present in a cluster. 
The secondary particles 
are expected to be more abundant in the denser central 
parts of clusters, where the magnetic field is also larger
and a correspondingly large contribution to the radio 
brightness is expected. In the outskirts of the cluster,
secondary electrons/positrons are more sparse because of the lower
gas density and a dominant primary component should emerge.
Therefore the phenomenology of the non thermal activity
in a cluster may reasonably be expected to be quite complex.
Contrary to the {\it classical} models of reacceleration 
of relic primary electrons, the models discussed in this
paper do predict that clusters may also be sources of 
gamma rays both due to the decay of neutral pions and
to radiative processes of secondary electrons and positrons, whose
spectrum at energies above those at which a reacceleration
bump is generated is left basically unchanged. 

We applied our calculations to a Coma-like cluster in order to 
check if the main phenomenological aspects of the non--thermal 
activity may be reproduced. 
In doing that we are forced to make the simple assumption
that MHD turbulence 
uniformly fills the cluster volume and that relativistic
particles are efficiently mixed with the thermal pool.
Clearly more detailed calculations are desirable, possibly 
making use of next generation numerical simulations.

At variance with standard reacceleration models, in which the source
of relic electrons is essentially a free parameter, 
here the injection of secondary particles is self--regulated by the 
relativistic hadrons and by the reacceleration process
itself, and this increases the predictive power of the model.
We find that the synchrotron emissivity at 330 MHz from the core region
allows us to infer useful bounds on the energy content in the
form of relativistic protons: in general, as illustrated in 
Fig. 4, observations are hardly explained unless the energy 
density in cosmic rays at the beginning of the acceleration 
stage is typically less than about $0.5\%$ of the thermal 
energy density in the cluster center (at least when both the 
central brightness and the relatively flat spectral index, 
$\alpha \sim 0.7-0.9$, of the core of Coma C are considered).
The rate of injection of energy in the form of Alfv\'en 
waves is effectively constrained and depends on the ratio
${\cal E}_p/{\cal E}_{th}$.
For ${\cal E}_p/{\cal E}_{th}$ larger than a few times $10^{-4}$ 
the injection rate is required to be of the order of $\sim 10^{-28}
\rm erg~s^{-1}~cm^{-3}$ (roughly $2\%$ of the thermal 
energy during the reacceleration time). A larger rate implies an 
even lower allowed fraction of cosmic rays in the ICM, while
an injection rate significantly smaller would produce a synchrotron
spectrum steeper than that observed. Clearly the limits are more 
stringent when the magnetic field is larger.
Similarly these limits are expected to be more stringent
when the real injected spectrum of cosmic ray protons 
is considerably steeper than $s=2.2$ (adopted throughout the 
paper) since in this case the back--reaction of protons
on the waves is even stronger (Paper I).
Finally, for acceleration periods longer than a few $10^8$yrs
(adopted in Sect.~5) the limits are expected to be slightly
more stringent since in this case 
the energy budget of protons increases, while the limits on
${\cal E}_p/{\cal E}_{th}$ would be slightly less stringent for
shorter reacceleration stages (provided that they are longer
than the typical acceleration time).

We studied in detail how the volume integrated synchrotron spectrum 
and the radial profile of the radio emission change when the 
contribution of reaccelerated secondary electrons and positrons 
is taken into
account. The dependence of the results upon the choice of the 
several parameters involved is illustrated and summarized in 
Sections 5.3 and 5.4. In principle given viable assumptions
the radio emission of Coma 
(volume integrated spectrum and radial profile) could be explained 
in terms of reaccelerated secondary electrons and positrons only. 
In particular, we have shown that the reacceleration of secondary 
leptons could produce synchrotron profiles which are broader 
than those expected from the 
standar acceleration model, although the broadening 
is still not sufficient to explain the most extended
radio halos (at least under the assumptions in Sect.~5).
On the other hand the observed, though controversial hard 
X-ray excess does require an additional component, that can 
plausibly be associated with primary electrons in the outskirts
of the cluster, where secondary particles are not abundant enough.

The possibility to demonstrate the existence
of cosmic ray hadrons and secondary electrons/positrons 
in the ICM is related to
the detection of gamma rays due to the decay of the neutral
pions and of circular polarization (provided this is not 
shrouded by the effect of Faraday depolarization in the ICM).
In this respect the {\it Hybrid Models} presented here are
qualitatively similar to the secondary models.
On the other hand, one way of testing a scenario in which non 
thermal radiation is
generated by reaccelerated leptons of both primary and 
secondary origin is to look for a radio tail that should 
be produced through synchrotron emission of those secondary 
leptons that are not affected by resonant interactions 
with Alfv\'en waves (namely with $\gamma>10^5$). This radio 
flux should appear as a sort of recovery of the radio emissivity
at high frequency, above the cutoff typical of the reacceleration
scenarios. This flux level might be accessible to next generation
radio telescopes, such as SKA, provided the radio flux at such 
high frequencies is not overwhelmingly smaller than the microwave
flux at the same frequecnies.

In this paper we focus on a specific aspect that is
the effect of MHD turbulence on the acceleration of both
primary and secondary relativistic
particles in the ICM and on the related non--thermal emission
from galaxy clusters.
If future observations of radio halos will confirm the presence 
of distinctive features (for example spectral steepenings)
which are expected in the case of {\it in situ} 
acceleration of the emitting particles, MHD turbulence in
the ICM would be unavoidable.
In this case, detailed calculations of particle acceleration
will be important 
to indirectly constrain the presence and the properties of the MHD turbulence
in the ICM and hopefully its connection
with the magnetic field amplification.

In general, turbulence could play a role in several aspects of the 
physics of the ICM.  
Large-scale turbulent motions in the ICM may provide a substantial 
pressure support to the ICM (Kulsrud et al. 1997;
Roettiger et al. 1997; Ricker \& Sarazin 2001), and
in addition to other proposed mechanisms (e.g.,
B\"ohringer et al., 2002; Ciotti \& Ostriker, 2001),
the dissipation of turbulent energy can provide a source of
heating to balance the cooling of cluster cores
(Fujita, Matsumoto \& Wada 2004).
The knowledge of the basic aspects of MHD turbulence in galaxy
clusters is also crucial to model the transport of heat and
metals in the ICM (Cho et al., 2003; Voigt \& Fabian 2004).

Future experiments, such as ASTRO-E2
(and NEXT), would hopefully constrain the energy budget associated to
the turbulent eddies in the ICM by looking at the profile of the
FeK--lines (and other) in the X--ray spectrum of galaxy clusters
(Sunyaev et al., 2003).

\section{Acknowledgements}
We acknowledge L.Feretti and the anonymous referee for 
valuable comments on the presentation of the paper.
We acknowledge partial support from MIUR 
through grant PRIN2004.
GB acknowledge partial support from INAF through grant D4/03/15.

\end{document}